\documentclass[12pt,a4paper]{article}
\pdfoutput=1
\usepackage[utf8]{inputenc}
\usepackage[table,dvipsnames]{xcolor}
\usepackage{epsf,amsmath,amssymb,graphicx,dcolumn}
\usepackage{caption}
\usepackage{scalefnt,ulem,pstricks}
\usepackage{booktabs,multirow,tabularx}
\usepackage[colorlinks=true,allcolors={blue!70!black}]{hyperref}
\usepackage{cleveref}
\usepackage{rotating}
\usepackage{microtype}
\usepackage{makecell}
\usepackage{cancel}
\usepackage{afterpage}
\usepackage[titletoc,title]{appendix}
\usepackage[numbers,sort&compress]{natbib}
\usepackage{amsmath,amsfonts,amsthm,bm}
\usepackage[tikz]{bclogo}
\usepackage{subcaption}
\usepackage{tikz-feynman}
\providecommand{\href}[2]{#2}

\newcommand\F{${\rm F}$}
\newcommand\FJ{${\rm FJ}$}
\newcommand\FJJ{${\rm FJJ}$}

\newcommand{\pt}{{p_{\text{\scalefont{0.77}T}}}}

\newcommand{\ptrad}{{p_{\text{\scalefont{0.77}T,rad}}}}

\newcommand{\mz}{{m_{\text{\scalefont{0.77}Z}}}}

\newcommand{\mt}{{m_{\text{\scalefont{0.77}t}}}}

\newcommand{\noun}[1]{{\scshape #1}}

\newcommand{\POWHEG}{\noun{Powheg}}

\newcommand{\POWHEGBOXRES}{\noun{Powheg-Box-Res}}

\newcommand{\minlo}{{\noun{MiNLO$^{\prime}$}}}
\newcommand{\minnlo}{{\noun{MiNNLO$_{\rm PS}$}}}
\newcommand{\Matrix}{{\noun{Matrix}}}
\newcommand{\OpenLoops}{{\noun{OpenLoops}}}
\newcommand{\PYTHIA}[1]{\noun{Pythia{#1}}}

\usepackage{xcolor}

\newcommand{\tmop}[1]{\ensuremath{\operatorname{#1}}}

\newcommand\as{\alpha_{\mathrm{S}}}

\def\to{\rightarrow}

\def\mt{m_t}

\def\mz{m_Z}

\newcommand{\eqn}[1]{Eq.\,(\ref{#1})}

\newcommand{\fig}[1]{Figure~\ref{#1}}

\newcommand{\tab}[1]{Table~\ref{#1}}
\newcommand{\sct}[1]{Section~\ref{#1}}

\def\citere#1{\mbox{Ref.~\cite{#1}}}
\def\citeres#1{\mbox{Refs.~\cite{#1}}}

\newcommand{\abbrev}{}

\newcommand{\nnlo}{\text{\abbrev NNLO}}

\usepackage{etoolbox}
\makeatletter
\patchcmd{\@sect}{#8}{\boldmath #8}{}{}
\let\ori@chapter\@chapter
\def\@chapter[#1]#2{\ori@chapter[\boldmath#1]{\boldmath#2}}
\makeatother

\begin{document} 
\begin{flushright}
\vspace*{-1.5cm}
MPP-2024-120
\end{flushright}
\vspace{0.cm}

\begin{center}
{\Large \bf Higgs-boson production in the full theory at NNLO+PS}
\end{center}

\begin{center}
  {\bf Marco Niggetiedt} and {\bf Marius Wiesemann}

Max-Planck-Institut f\"ur Physik, Boltzmannstraße 8, 85748 Garching, Germany

\href{mailto:marco.niggetiedt@mpp.mpg.de}{\tt marco.niggetiedt@mpp.mpg.de},
\href{mailto:marius.wiesemann@mpp.mpg.de}{\tt marius.wiesemann@mpp.mpg.de}

\end{center}

\begin{center} {\bf Abstract} \end{center}\vspace{-1cm}
\begin{quote}
\pretolerance 10000

We consider the production of a Standard-Model (SM) Higgs boson in gluon fusion
in hadronic collisions and compute the QCD corrections up to next-to-next-to-leading order (NNLO) and match them to parton showers (NNLO+PS). 
The complete dependence on the top-quark mass is taken into account
without making any approximations to the top quark loops mediating the 
coupling between the gluons and the Higgs boson. To this end, we have included
the $gg\to H$ amplitudes up to three loops and the $pp\to H$+jet amplitudes up 
to two loops in the full SM theory.
This is the first fully differential calculation of the top-quark mass effects up to NNLO in QCD, 
and we study their impact on relevant observables for the LHC.

\end{quote}

\parskip = 1.2ex

\section{Introduction}
\label{sec:intro}

Measurements of the Higgs boson and its properties are one of the cornerstones of the rich 
physics programme at the Large Hadron Collider (LHC). After its discovery a decade ago~\cite{ATLAS:2012yve,CMS:2012qbp}, the characterization of the Higgs boson has become 
one of the major quests of the particle physics community.
Being naturally the least explored sector in the Standard Model (SM), 
the Higgs boson offers a large potential for finding discrepancies with the SM 
in the search for new-physics phenomena.
Nevertheless, so far, the measured Higgs couplings to top ($t$) and bottom ($b$) quarks, to $W$ and $Z$ bosons, and to $\tau$ leptons draw a picture fully consistent with the SM expectation \cite{ATLAS:2022vkf,CMS:2022dwd}.
With the continuously increasing data taking at the LHC, however, the accurate extraction of 
Higgs properties provides a prime candidate to discover small deviations from SM predictions.
To maximize our chances to find such deviation does not only require to reduce the experimental 
uncertainties through high-statistics measurements, but also to push forward theoretical simulations
of LHC collisions to the highest possible precision.

In the SM, Higgs-boson production in hadronic collisions proceeds predominantly through gluon fusion, 
where the Higgs--gluon coupling is mediated by a heavy-quark loop, with the top quark providing the largest 
contribution. The most accurate calculations for the gluon-fusion process so far have been done 
in the approximation of an infinitely heavy top quark, referred to as heavy top limit (HTL).
The total inclusive cross section at next-to-next-to-leading order (NNLO) in QCD has been obtained 
in this approximation already two decades ago \cite{Harlander:2002wh,Anastasiou:2002yz,Ravindran:2003um} and even the calculation 
of the next-to-NNLO (N$^3$LO) cross section lies quite a few years back \cite{Anastasiou:2015vya,Mistlberger:2018etf}. Also several 
fully-differential calculations in the HTL appeared over the years: at NNLO QCD accuracy for $H$ \cite{Anastasiou:2004xq,Catani:2007vq}
and $H$+jet production \cite{Boughezal:2013uia,Chen:2014gva,Boughezal:2015dra,Chen:2016zka}, 
at NNLO QCD matched to parton showers (NNLO+PS) \cite{Hamilton:2013fea,Hoche:2014dla,Monni:2019whf,Monni:2020nks}, and recently even at N$^3$LO QCD \cite{Cieri:2018oms,Dulat:2018bfe,Chen:2021isd,Billis:2021ecs}.
As far as the quark-mass effects missing in the HTL approximation are concerned, there have been various studies to assess the impact of both the top and the bottom mass in the past at the level of the fully inclusive 
cross section, see \citeres{Marzani:2008az,Harlander:2009mq,Harlander:2009my,Pak:2009dg,Pak:2011hs}, and at the level of the fully differential cross section, see \citeres{Spira:1995rr,Harlander:2005rq,Bagnaschi:2011tu,Harlander:2012hf,Mantler:2012bj,Banfi:2013eda,Grazzini:2013mca,Harlander:2014uea,Neumann:2014nha,Hamilton:2015nsa,Mantler:2015vba,Bagnaschi:2015qta,Bagnaschi:2015bop,Frederix:2016cnl,Neumann:2016dny,Melnikov:2016emg,Caola:2016upw,Lindert:2017pky,Liu:2017vkm,Jones:2018hbb,Caola:2018zye,Chen:2021azt}.
However, it was only recently that the full dependence on the top-quark mass \cite{Czakon:2021yub} and on the top-bottom interference \cite{Czakon:2023kqm} was 
calculated for the total inclusive cross section at NNLO QCD.

In this paper, we present the first differential calculation of NNLO QCD corrections in the full theory, i.e.\ without 
any approximations for the top quark mediating the Higgs--gluon interaction, and match them consistently
to the parton shower. To this end we employ the \minnlo{} formalism to implement an NNLO+PS generator
for the Higgs-boson production in gluon fusion. Being a loop induced process, this calculation entails
the full three-loop amplitudes for $pp\to H$ and the full two-loop amplitudes for $gg\to H$+jet production.
We study in detail the impact of the complete top-quark mass dependence and assess the quality of previously 
employed approximations for the top-quark mass effects at NNLO through comparison to our full-theory results.

This letter is organized as follows: We start by outlining the calculation in \sct{sec:outline} before we discuss the
computation of the higher-loop amplitudes in \sct{sec:loopamps}. \sct{sec:approx} describes 
various approximations of the top-mass effects that we use in comparison to our full calculation.
\sct{sec:minnlo} contains a brief summary of the \minnlo{} method.
Phenomenological results are presented in \sct{sec:results}, and we summarize in \sct{sec:summary}.

\section{Outline of the calculation}
\label{sec:outline}

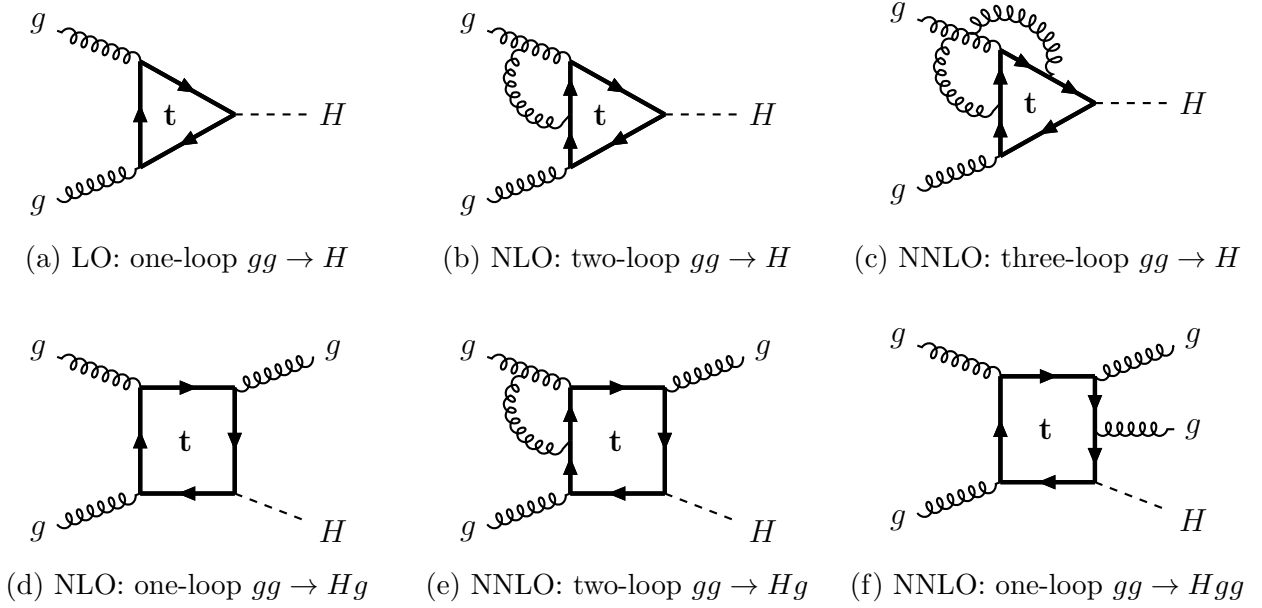
\begin{figure}[t]
  \begin{center}
    \begin{subfigure}[b]{.3\linewidth}
      \centering
\begin{tikzpicture}
  \begin{feynman}
	\vertex (a1) at (-0.1,0.1) {\( g\)};
	\vertex (a2) at (-0.1,-2.3) {\( g\)};
	\vertex (a3) at (1.25,-0.4);
	\vertex (a4) at (1.25,-1.8);
	\vertex (a5) at (2.5,-1.1);
	\vertex (a6) at (3.8,-1.1){\( H\)};
	\vertex (a7) at (1.65,-1.1){\(\bf{}t\)};
        \diagram* {
          {[edges=fermion]
            (a5)--[fermion, ultra thick](a4)--[fermion, ultra thick](a3)--[fermion, ultra thick](a5),
          },
          (a5)--[scalar, thick](a6),
          (a2)--[gluon,thick](a4),
          (a3)--[gluon,thick](a1),
        };
  \end{feynman}
\end{tikzpicture}
\caption{LO: one-loop $gg\to H$}
        \label{subfig:LO}
\end{subfigure}%
\hspace{0.3cm}
\begin{subfigure}[b]{.3\linewidth}
  \centering
\begin{tikzpicture}
  \begin{feynman}
	\vertex (a1) at (-0.1,0.1) {\( g\)};
	\vertex (a12) at (0.65,-0.275);
	\vertex (a2) at (-0.1,-2.3) {\( g\)};
	\vertex (a3) at (1.25,-0.4);
	\vertex (a32) at (1.25,-1.1);
	\vertex (a4) at (1.25,-1.8);
	\vertex (a5) at (2.5,-1.1);
	\vertex (a6) at (3.8,-1.1){\( H\)};
	\vertex (a7) at (1.65,-1.1){\(\bf{}t\)};
        \diagram* {
          {[edges=fermion]
            (a5)--[fermion, ultra thick](a4)--[fermion, ultra thick](a32)--[fermion, ultra thick](a3)--[fermion, ultra thick](a5),
          },
          (a5)--[scalar, thick](a6),
          (a2)--[gluon,thick](a4),
          (a12)--[gluon,thick,half right](a32),
          (a3)--[gluon,thick](a1),
        };
  \end{feynman}
\end{tikzpicture}
\caption{NLO: two-loop $gg\to H$}
        \label{subfig:gg}
\end{subfigure}%
\hspace{0.3cm}
\begin{subfigure}[b]{.3\linewidth}
  \centering
\begin{tikzpicture}
  \begin{feynman}
	\vertex (a1) at (-0.1,0.1) {\( g\)};
	\vertex (a12) at (0.65,-0.275);
	\vertex (a13) at (0.85,-0.18);
	\vertex (a2) at (-0.1,-2.3) {\( g\)};
	\vertex (a3) at (1.25,-0.4);
	\vertex (a32) at (1.25,-1.1);
	\vertex (a4) at (1.25,-1.8);
	\vertex (a5) at (2.5,-1.1);
	\vertex (a52) at (1.875,-0.75);
	\vertex (a6) at (3.8,-1.1){\( H\)};
	\vertex (a7) at (1.65,-1.1){\(\bf{}t\)};
        \diagram* {
          {[edges=fermion]
            (a5)--[fermion, ultra thick](a4)--[fermion, ultra thick](a32)--[fermion, ultra thick](a3)--[fermion, ultra thick](a52)--[fermion, ultra thick](a5),
          },
          (a5)--[scalar, thick](a6),
          (a2)--[gluon,thick](a4),
          (a12)--[gluon,thick,half right](a32),
          (a13)--[gluon,thick,half left](a52),
          (a3)--[gluon,thick](a1),
        };
  \end{feynman}
\end{tikzpicture}\vspace{0.15cm}
\caption{NNLO: three-loop $gg\to H$}
        \label{subfig:gg}
\end{subfigure}\\[0.5cm]
    \begin{subfigure}[b]{.3\linewidth}
      \centering
\begin{tikzpicture}
  \begin{feynman}
	\vertex (a1) at (-0.1,0.1) {\( g\)};
	\vertex (a2) at (-0.1,-2.3) {\( g\)};
	\vertex (a3) at (1.25,-0.4);
	\vertex (a4) at (1.25,-1.8);
	\vertex (a5) at (2.5,-0.4);
	\vertex (a55) at (2.5,-1.8);
	\vertex (a6) at (3.8,-2.3){\( H\)};
	\vertex (a7) at (1.85,-1.1){\(\bf{}t\)};
	\vertex (a8) at (3.8,0.1) {\( g\)};
        \diagram* {
          {[edges=fermion]
            (a55)--[fermion, ultra thick](a4)--[fermion, ultra thick](a3)--[fermion, ultra thick](a5)--[fermion, ultra thick](a55),
          },
          (a55)--[scalar, thick](a6),
          (a2)--[gluon,thick](a4),
          (a3)--[gluon,thick](a1),
          (a5)--[gluon,thick](a8),
        };
  \end{feynman}
\end{tikzpicture}
\caption{NLO: one-loop $gg\to Hg$}
        \label{subfig:qq}
\end{subfigure}%
\hspace{0.3cm}
\begin{subfigure}[b]{.3\linewidth}
  \centering
\begin{tikzpicture}
  \begin{feynman}
	\vertex (a1) at (-0.1,0.1) {\( g\)};
	\vertex (a12) at (0.65,-0.275);
	\vertex (a2) at (-0.1,-2.3) {\( g\)};
	\vertex (a3) at (1.25,-0.4);
	\vertex (a32) at (1.25,-1.1);
	\vertex (a4) at (1.25,-1.8);
	\vertex (a5) at (2.5,-0.4);
	\vertex (a55) at (2.5,-1.8);
	\vertex (a6) at (3.8,-2.3){\( H\)};
	\vertex (a7) at (1.85,-1.1){\(\bf{}t\)};
	\vertex (a8) at (3.8,0.1) {\( g\)};
        \diagram* {
          {[edges=fermion]
            (a55)--[fermion, ultra thick](a4)--[fermion, ultra thick](a32)--[fermion, ultra thick](a3)--[fermion, ultra thick](a5)--[fermion, ultra thick](a55),
          },
          (a55)--[scalar, thick](a6),
          (a2)--[gluon,thick](a4),
          (a12)--[gluon,thick,half right](a32),
          (a3)--[gluon,thick](a1),
          (a5)--[gluon,thick](a8),
        };
  \end{feynman}
\end{tikzpicture}
\caption{NNLO: two-loop $gg\to Hg$}
        \label{subfig:gg}
\end{subfigure}%
\hspace{0.3cm}
\begin{subfigure}[b]{.3\linewidth}
  \centering
\begin{tikzpicture}
  \begin{feynman}
	\vertex (a1) at (-0.1,0.1) {\( g\)};
	\vertex (a2) at (-0.1,-2.3) {\( g\)};
	\vertex (a3) at (1.25,-0.4);
	\vertex (a4) at (1.25,-1.8);
	\vertex (a5) at (2.5,-0.4);
	\vertex (a52) at (2.5,-1.1);
	\vertex (a522) at (3.8,-1.1){\( g\)};
	\vertex (a55) at (2.5,-1.8);
	\vertex (a6) at (3.8,-2.3){\( H\)};
	\vertex (a7) at (1.85,-1.1){\(\bf{}t\)};
	\vertex (a8) at (3.8,0.1) {\( g\)};
        \diagram* {
          {[edges=fermion]
            (a55)--[fermion, ultra thick](a4)--[fermion, ultra thick](a3)--[fermion, ultra thick](a5)--[fermion, ultra thick](a52)--[fermion, ultra thick](a55),
          },
          (a55)--[scalar, thick](a6),
          (a2)--[gluon,thick](a4),
          (a3)--[gluon,thick](a1),
          (a5)--[gluon,thick](a8),
          (a52)--[gluon,thick](a522),
        };
  \end{feynman}
\end{tikzpicture}\vspace{0.15cm}
\caption{NNLO: one-loop $gg\to Hgg$}
        \label{subfig:gg}
\end{subfigure}\\
\end{center}
\caption{\label{fig:diagrams} Sample Feynman diagrams entering  
  $pp\to H$ production up to NNLO.}
\end{figure}

We consider the process
\begin{align}
pp\to H + X\,,
\end{align}
inclusive over the radiation of any extra particles $X$, and we compute
radiative corrections to NNLO QCD in perturbation theory. 
At LO, Higgs-boson production proceeds via the fusion of two gluons
and the Higgs--gluon coupling is induced by a quark loop. The dominant 
contribution comes from the top quark due to its large Yukawa interaction, see \fig{fig:diagrams}\,(a)
for the corresponding Feynman diagram.
At NLO, the virtual contribution is computed from the two-loop $gg\to H$ amplitude, see \fig{fig:diagrams}\,(b),
and the real contribution corresponds to a one-loop diagram for Higgs production with an extra parton 
in the final state, i.e.\ the process $pp\to Hj$, see \fig{fig:diagrams}\,(d) for an example.
At NNLO, the computation of the three-loop $gg\to H$ amplitude is required, see \fig{fig:diagrams}\,(c), the
real-virtual contribution corresponding to the two-loop $pp\to Hj$ amplitude enters, see \fig{fig:diagrams}\,(e), 
and the double-real correction has to be included through the one-loop amplitude for Higgs production with two 
extra partons, i.e.\ the process $pp\to Hjj$, see \fig{fig:diagrams}\,(f) for an example.
Using the \minnlo{} method \cite{Monni:2019whf,Monni:2020nks}, which will be sketched below, we combine the separately divergent amplitudes to 
compute the NNLO QCD cross section and perform its matching to a parton shower.
Note that beyond LO, also diagrams with quarks in the initial state enter the calculation of the cross section,
which we do not explicitly depict here.

Our \minnlo{} $H$ generator in the full theory is implemented
 in the \POWHEGBOXRES{} framework \cite{Jezo:2015aia}.
The one-loop-squared amplitudes for the $pp\to H$, $pp\to Hj$, and $pp\to Hjj$ processes are computed through 
\OpenLoops{}~\cite{Cascioli:2011va,Buccioni:2017yxi,Buccioni:2019sur}. To this end, 
we have extended the \OpenLoops{} interface in \POWHEGBOXRES{} developed in \citere{Jezo:2016ujg}
to loop-induced processes. The calculation of the employed two-loop and three-loop amplitudes
has been performed in \citeres{Czakon:2020vql,Czakon:2021yub}, as described below. The two-loop and three-loop $gg\to H$ amplitudes 
are evaluated efficiently through a deep asymptotic expansion in the low-energy limit. The calculation of the two-loop $pp\to Hj$ contribution 
is numerically quite intensive and we have derived a dense grid in the phase space and implemented 
a numerical interpolation based on cubic b-splines. We have checked the validity 
of the interpolation by performing the same interpolation also for the corresponding one-loop $pp\to Hj$ 
amplitude in the HTL by comparing it to the exact result.
We provide more details on the calculation of these amplitudes in the following section.

\section{Two-loop and three-loop amplitudes}
\label{sec:loopamps}

As far as the determination of NNLO corrections including the exact top-quark mass dependence is concerned, the efficient evaluation of the two-loop corrections to the $pp\to Hj$ process constitutes the main obstacle of the computation due to the rich analytic structure of the underlying Feynman integrals. The workflow of the corresponding calculation has been outlined in \citere{Czakon:2021yub}. First, the amplitudes have been reduced to linear combinations of master integrals with the public software \texttt{Kira}$\oplus$\texttt{FireFly} \cite{Maierhofer:2017gsa,Maierhofer:2018gpa,Klappert:2020nbg,Klappert:2019emp,Klappert:2020aqs}. A particular basis of master integrals suitable for numerical evaluation has been chosen based on the differential equations \cite{Kotikov:1990kg,Kotikov:1991pm,Kotikov:1991hm,Remiddi:1997ny}, which govern the evolution of the set of master integrals in the space of kinematic invariants and masses. In order to employ the differential equations, an initial condition must be provided, which we obtain from a diagrammatic large-mass expansion \cite{Gorishnii:1989dd,Smirnov:1990rz,Smirnov:1994tg,Smirnov:2002pj}. It is worth to note that the analytic calculation of the relevant master integrals has previously been addressed in the literature \cite{Bonciani:2016qxi,Bonciani:2019jyb,Frellesvig:2019byn}. However, even for a subset of planar master integrals the appearance of elliptic sectors has been observed, significantly increasing the complexity of the analytic evaluation. Therefore, we solved the complete set of master integrals numerically in our own framework. 

Since the determination of top-quark mass effects is at the core of the study at hand, we fix the ratio between the mass of the Higgs-boson and the mass of the top-quark according to $m_t^2/m_H^2=23/12$, thereby reducing the dimension of the physical parameter space. Without imposing any constraints on the kinematic variables and masses, the parameter space for the two-loop $pp\to Hj$ amplitudes is spanned by three dimensionless variables (build from the overall four independent dimensionful variables):

\begin{equation}
\rho=\frac{m_t^2}{\hat{s}}\,, \qquad z=1-\frac{m_H^2}{\hat{s}}\,, \qquad \lambda=\frac{\hat{t}}{\hat{t}+\hat{u}}\,,
\end{equation}

with the standard definition of the Mandelstam variables, $\hat{s}$, $\hat{t}$, and $\hat{u}$. The two-dimensional subspace parametrized in terms of $z\in (0,1)$ and $\lambda\in (0,1)$ is probed using the differential equations for the master integrals. More precisely, employing the differential equations in $\rho$, the initial condition located at $\rho\to\infty$ is transported to the two-dimensional plane defined by $m_t^2/m_H^2=const$. Subsequently, the endpoint of the previous evolution is used as a seed for further evolutions in the directions of $z$ and $\lambda$ to sample the physical phase space. 

It is important to point out that only the phase space region above the top-quark pair threshold needs to be sampled by means of numerical evolutions. For the remaining region including the soft region, we exploit the aforementioned large-mass expansion, which we derived to order $(1/\rho)^{40}$, guaranteeing an efficient evaluation of the amplitudes. We note that the leading term of the expansion corresponds to the HTL and serves as a cross-check for the correctness of our amplitudes. 

With the aid of the deep large-mass expansions and the differential equations for numerical evolutions in the $z$-$\lambda$-plane for fixed $m_t^2/m_H^2$, we generated dense regular two-dimensional grids in $z$ and $\lambda$, which are suitable for the interpolation of the relevant amplitudes $gg\to Hg$ and $q\bar{q}\to Hg$ and the permutations of external partons of the latter one. Every grid consists of approximately $7 \times 10^6$ high-precision samples of the amplitudes. The grids extend far into the soft and collinear limits, so that no extrapolation is necessary for our purposes. We anticipate that the same grids can also be used in phenomenological applications for slightly different fixed ratios $m_t^2/m_H^2$ deviating by no more than 1\% from the actual ratio without spoiling the outcome of the study. The interpolation is performed through cubic b-splines using the {\sc bspline-fortran} 
code\footnote{See code under \url{https://github.com/jacobwilliams/bspline-fortran} developed by Jacob Williams.}. The interpolation procedure
was validated at the level of the HTL and at the level of the Born amplitude, since for each of them a fast exact implementation exists.

We note that the technology used to obtain the amplitudes for the top quark with a fixed ratio $m_t^2/m_H^2$ can also be extended to light quark masses.
However, due to the smallness of the masses of the bottom and charm quark, only the exact numerical solution of the differential equations is viable, and
 the corresponding calculation is  more involved with regard to the treatment of numerical instabilities. 
We plan to implement additional grids for the $pp\to H$+jet amplitudes at different  ratios $m_q^2/m_H^2$ in future work, specifically for different values of the top-quark, bottom-quark, and charm-quark masses. 
Furthermore, we reckon that the interpolation of the amplitudes with a fixed mass ratio can subsequently be extended through an additional interpolation in the $m_q^2/m_H^2$ ratio, 
hereby relaxing the constraint on the quark mass completely. 

Regarding the three-loop double-virtual corrections, the corresponding form factor parametrizing the amplitude for the production of a possibly off-shell Higgs-boson via the scattering of a pair of gluons is available for arbitrary quark flavors circulating in the loops \cite{Czakon:2020vql,Niggetiedt:2023uyk}. The form factor is expressed in terms of expansions around special kinematic points and is supplemented by high-precision numerical samples. It is sufficient for our application to employ the provided deep asymptotic expansion of the form factor in the low-energy limit as it covers the mass-effects of the top-quark. 

\section{Approximations of the top-mass effects}
\label{sec:approx}

Apart from the full-theory implementation retaining the complete dependence on the top-quark loop, 
we have implemented various approximations for the top quark, which allows us to determine the
relevance of the full top-mass effects compared to approximations previously employed in the literature.
First of all, we have included the HTL by assuming an infinitely heavy top quark and integrating out 
the top quark. As a result, in the HTL Feynman diagrams, which due to their simplicity we do not give here explicitly, 
the top-quark loop in \fig{fig:diagrams} effectively shrinks to a point, inducing a point-like Higgs--gluon vertex in this effective field theory. 
Therefore, all HTL amplitudes are substantially simpler to calculate as the number of (massive) loops and the number of scales reduces by one. Indeed, the LO diagram
in the HTL becomes a tree-level $2\to 1$ diagram where the Higgs directly couples to the gluons. For these reasons
the HTL has been employed extensively in the past to obtain radiative corrections to Higgs production in gluon fusion,
with the most recent advancement being the N$^3$LO corrections in the approximation of an infinitely heavy top quark \cite{Anastasiou:2015vya,Mistlberger:2018etf}.

Between the most crude approximation of the HTL and the full theory computation there are several intermediate 
approximations that can be taken.
Firstly, the radiative corrections computed in the HTL can be rescaled by the LO $gg\to H$ cross sections
in the full theory (FT), which we shall refer to as HEFT in what follows:
\begin{align}
{\mathrm d}\sigma_{\rm HEFT}^{\rm NNLO} = {\mathrm d}\sigma_{\rm FT}^{\rm LO} \cdot {\mathrm d}\sigma_{\rm HTL}^{\rm NNLO}/{\mathrm d}\sigma_{\rm HTL}^{\rm LO}\,.\label{eq:HEFT}
\end{align}
This approximation, however, only affects the normalization. Differential observables sensitive to a jet in the final state, such as the transverse momentum
of the Higgs boson or of the jet, will be described by the HTL amplitude, which is not a good approximation, especially at large transverse momentum.
Therefore, we have implemented further approximations of the FT. In all those approximations we keep the one-loop squared Born-level amplitudes for 
$pp\to H$, $pp\to Hj$, and $pp\to Hjj$ exact, while the two-loop $pp\to Hj$ amplitudes, which are the most complicated
part of our calculation, are computed in the HTL and rescaled differentially by the full $pp\to Hj$ LO amplitude. We can write this at the level of the finite remainder
$|\mathcal{R}_{Hj}\rangle$ of the $pp\to Hj$ amplitudes as follows:\footnote{We define the finite remainder as described in \citeres{Becher:2009cu,Becher:2009qa}. Note that this definition should not 
affect the given approximation.}
\begin{align}
\label{eq:Hjapprox}
2\,{\rm Re}\langle \mathcal{R}_{Hj}^{{(0)}} |\mathcal{R}_{Hj}^{{(1)}}\rangle_{\rm FT} \approx 2\,{\rm Re}\langle\mathcal{R}^{(0)}_{Hj} |\mathcal{R}^{(1)}_{Hj}\rangle_{\rm HTL} \cdot \langle\mathcal{R}^{(0)}_{Hj} |\mathcal{R}^{(0)}_{Hj}\rangle_{\rm FT}\,/\,\langle\mathcal{R}^{(0)}_{Hj} |\mathcal{R}^{(0)}_{Hj}\rangle_{\rm HTL}\,,
\end{align}
where we recall that in the FT $\mathcal{R}_{Hj}^{{(0)}}$ and $\mathcal{R}_{Hj}^{{(1)}}$ correspond to diagrams at one-loop and two-loop, respectively, while in the HTL they are obtained from tree-level and one-loop amplitudes, respectively.
Additionally, we include three possibilities to approximate the two-loop and three-loop $gg\to H$ amplitudes. In the first approximation, denoted as FT-approx-1, we apply \eqn{eq:Hjapprox}
and in addition:
\begin{align}
\label{eq:H1}
2\,{\rm Re}\langle \mathcal{R}_{H}^{{(0)}} |\mathcal{R}_{H}^{{(1)}}\rangle_{\rm FT} \approx 2\,{\rm Re}\langle\mathcal{R}^{(0)}_{H} |\mathcal{R}^{(1)}_{H}\rangle_{\rm HTL} \cdot \langle\mathcal{R}^{(0)}_{H} |\mathcal{R}^{(0)}_{H}\rangle_{\rm FT}\,/\,\langle\mathcal{R}^{(0)}_{H} |\mathcal{R}^{(0)}_{H}\rangle_{\rm HTL}\,,
\end{align}
and
\begin{align}
\begin{split}
\label{eq:H2}
\langle \mathcal{R}_{H}^{{(1)}} |\mathcal{R}_{H}^{{(1)}}\rangle_{\rm FT} &\approx \langle\mathcal{R}^{(1)}_{H} |\mathcal{R}^{(1)}_{H}\rangle_{\rm HTL} \cdot \langle\mathcal{R}^{(0)}_{H} |\mathcal{R}^{(0)}_{H}\rangle_{\rm FT}\,/\,\langle\mathcal{R}^{(0)}_{H} |\mathcal{R}^{(0)}_{H}\rangle_{\rm HTL}\,,\\
2\,{\rm Re}\langle \mathcal{R}_{H}^{{(0)}} |\mathcal{R}_{H}^{{(2)}}\rangle_{\rm FT} &\approx 2\,{\rm Re}\langle\mathcal{R}^{(0)}_{H} |\mathcal{R}^{(2)}_{H}\rangle_{\rm HTL} \cdot \langle\mathcal{R}^{(0)}_{H} |\mathcal{R}^{(0)}_{H}\rangle_{\rm FT}\,/\,\langle\mathcal{R}^{(0)}_{H} |\mathcal{R}^{(0)}_{H}\rangle_{\rm HTL}\,,
\end{split}
\end{align}
where $\mathcal{R}_{H}^{{(0)}}$, $\mathcal{R}_{H}^{{(1)}}$, and $\mathcal{R}_{H}^{{(2)}}$ are obtained from one-loop, two-loop, and three-loop amplitudes in the FT,
but only constitute to tree-level, one-loop, and two-loop computations in the HTL.
The second approximation, denoted as FT-approx-2, is exactly the same as FT-approx-1, but we drop the approximation of the two-loop $gg\to H$ amplitude in \eqn{eq:H1} 
and take it exact, given that it is known since many years ago \cite{Spira:1995rr}. We do exactly the same in our third approximation, denoted as FT-approx-3,
but instead of the approximation in \eqn{eq:H2} we use the full two-loop $gg\to H$ amplitude in the rescaling:
\begin{align}
\begin{split}
\label{eq:H22}
\langle \mathcal{R}_{H}^{{(1)}} |\mathcal{R}_{H}^{{(1)}}\rangle_{\rm FT} &\approx \langle\mathcal{R}^{(1)}_{H} |\mathcal{R}^{(1)}_{H}\rangle_{\rm HTL} \cdot 2\,{\rm Re}\langle\mathcal{R}^{(0)}_{H} |\mathcal{R}^{(1)}_{H}\rangle_{\rm FT}\,/\,2\,{\rm Re}\langle\mathcal{R}^{(0)}_{H} |\mathcal{R}^{(1)}_{H}\rangle_{\rm HTL}\,,\\
2\,{\rm Re}\langle \mathcal{R}_{H}^{{(0)}} |\mathcal{R}_{H}^{{(2)}}\rangle_{\rm FT} &\approx 2\,{\rm Re}\langle\mathcal{R}^{(0)}_{H} |\mathcal{R}^{(2)}_{H}\rangle_{\rm HTL} \cdot 2\,{\rm Re}\langle\mathcal{R}^{(0)}_{H} |\mathcal{R}^{(1)}_{H}\rangle_{\rm FT}\,/\,2\,{\rm Re}\langle\mathcal{R}^{(0)}_{H} |\mathcal{R}^{(1)}_{H}\rangle_{\rm HTL}\,.
\end{split}
\end{align}
Of course, there is a certain level of ambiguity involved, depending on how the finite remainders are defined in detail. However, 
FT-approx-1, FT-approx-2, FT-approx-3 are approximations that were feasible also without the complete computation in the FT,
and they allow us to determine the robustness of such approximations by comparing them to the full result. This can be useful
not only to validate legacy Higgs calculations that were applied previously, 
but also in view of applying such approximations beyond NNLO, and in particular to the production of a Higgs boson with higher multiplicities,
such as $pp\to Hjj$ production in gluon fusion.

\section{\minnlo{} method}
\label{sec:minnlo}

In the following, we provide a brief summary of the \minnlo{} method for colour-singlet production, which was originally
developed in \citeres{Monni:2019whf,Monni:2020nks} and has been applied to several processes by now \cite{Lombardi:2020wju,Lombardi:2021rvg,Buonocore:2021fnj,Lombardi:2021wug,Zanoli:2021iyp,Gavardi:2022ixt,Haisch:2022nwz,Lindert:2022qdd,Biello:2024vdh}. 
\minnlo{} is also the only NNLO+PS method that has been extended to processes with colour charges in initial and final state, namely
heavy-quark pair production \cite{Mazzitelli:2020jio,Mazzitelli:2021mmm,Mazzitelli:2023znt}, and 
very recently to the associated production of a heavy-quark pair and a colour singlet in \citere{Mazzitelli:2024ura}, where it was applied to $b\bar{b}Z$ production.

Starting from a \POWHEG{} \cite{Nason:2004rx,Nason:2006hfa,Frixione:2007vw,Alioli:2010xd} NLO+PS 
calculation for the production of a colour singlet (\F{}) with a jet, the \minnlo{} master formula can be expressed as
\begin{align}
\label{eq:master}
      {\rm d}\sigma_{\rm\scriptscriptstyle F}^{\rm MiNNLO_{PS}}={\rm d}\Phi_{\scriptscriptstyle\rm FJ}\,\bar{B}^{\,\rm MiNNLO_{\rm PS}}\,\times\,\left\{\Delta_{\rm pwg}(\Lambda_{\rm pwg})+
      {\rm d}\Phi_{\rm rad}\Delta_{\rm pwg}(\ptrad)\,\frac{R_{\scriptscriptstyle\rm FJ}}{B_{\scriptscriptstyle\rm FJ}}\right\}\,,
\end{align}
where the $\bar B$ function of \POWHEG{} is modified in such a way that it yields NNLO QCD
accuracy for the production of \F{} in the limit where QCD radiation becomes unresolved
\begin{align}
\label{eq:bbar}
\bar{B}^{\,\rm MiNNLO_{\rm PS}}= e^{-S}\,\left\{\frac{{\rm d}\sigma^{(1)}_{\scriptscriptstyle\rm FJ}}{{\rm d}\Phi_{\scriptscriptstyle\rm FJ}}\big(1+S^{(1)}\big)
+\frac{{\rm d} \sigma^{(2)}_{\scriptscriptstyle\rm FJ}}{{\rm d}\Phi_{\scriptscriptstyle\rm FJ}}+\left(D-D^{(1)}-D^{(2)}\right)\times F^{\rm corr}\right\}\,.
\end{align}
Here, $\Phi_{\scriptscriptstyle {\rm FJ}}$ denotes the \FJ{} phase space, $\Delta_{\rm pwg}$ is the \POWHEG{} Sudakov, and $\Phi_{\tmop{rad}} $ and $\ptrad$ are
the phase space and the transverse momentum of the second radiation. 
$B_{\scriptscriptstyle {\rm FJ}}$ and $R_{\scriptscriptstyle {\rm FJ}}$ are determined by the squared tree-level matrix elements for \FJ{} and \FJJ{} production, respectively.
In \eqn{eq:bbar}, ${\rm d}\sigma^{(1,2)}_{\scriptscriptstyle {\rm FJ}}$ denote the first- and second-order differential \FJ{} cross section. All other contributions originate
from the transverse-momentum ($\pt$) resummation formula, as explained in section 4 of \citere{Monni:2019whf},
\begin{align}
\label{eq:resum}
{\rm d}\sigma_{\scriptscriptstyle\rm F}^{\rm res}=\frac{{\rm d}}{{\rm
    d}\pt}\left\{e^{-S}\mathcal{L}\right\}=e^{-S}\underbrace{\left\{-S^\prime\mathcal{L}+\mathcal{L}^\prime\right\}}_{\equiv D}\,,
\end{align}
which defines the function $D$ in \eqn{eq:bbar}, and $e^{-S}$ is the Sudakov form factor, $S^{(1)}$ is the $\mathcal{O}(\as)$ term
in the expansion of its exponent, $\mathcal{L}$ denotes the luminosity factor up to NNLO,
which includes the convolution of the collinear coefficient functions with
the parton distribution functions (PDFs) and the squared hard-virtual
matrix elements for \F{} production. Note that within the \minnlo{} approach 
renormalization and factorization scales are set to $\pt$, except for the two overall
powers of $\alpha_s$, whose scale can be chosen arbitrarily.

The last term of the $\bar{B}$ function in \eqn{eq:bbar},
which is of order $\as^3(p_{\text{\scalefont{0.77}T}})$, adds the
relevant (singular) contributions necessary to reach NNLO accuracy
\cite{Monni:2019whf}, while regular contributions in $\pt$ are subleading at this order.
Note that, instead of truncating the singular contributions included through $D$ at $\as^3$, i.e.\ 
$\left(D-D^{(1)}-D^{(2)}\right)=D^{(3)}+\mathcal{O}(\as^4)$, as in the 
original \minnlo{} formulation of \citere{Monni:2019whf}, we follow the extension introduced in \citere{Monni:2020nks} 
and preserve the total derivative in \eqn{eq:resum}. This allows us to keep subleading logarithmic contributions beyond $\mathcal{O}(\as^3)$
that are relevant to achieve a better agreement with fixed-order NNLO results.
The factor $F^{\rm corr}$ in \eqn{eq:bbar} implements an appropriate functional form to spread 
$\left(D-D^{(1)}-D^{(2)}\right)$, which has Born-like kinematics, in $\Phi_{\scriptscriptstyle {\rm FJ}}$ 
within our event generator \cite{Monni:2019whf}.

Finally, the \minlo{} cross section is defined by simply dropping the last term of \eqn{eq:bbar}, which includes the NNLO corrections.
Thereby, one achieves a merging of $0$-jet and $1$-jet multiplicities at NLO QCD accuracy.

\section{Results}
\label{sec:results}
We present phenomenological results of our new \minnlo{} generator for Higgs production in gluon fusion in the full theory at the LHC with 13\,TeV centre-of-mass energy.
To study the effect of the full top-mass dependence in our calculation, we consider both the case of an on-shell Higgs boson and the one where the $H\to \gamma\gamma$ decay
is included in the zero-width approximation through the parton shower, with realistic fiducial cuts on the photons as applied by the experiments.

For the input parameters we set a Higgs mass of $m_H = 125$\,GeV. 
The top-quark is treated in the on-shell scheme with a mass of $\mt= 173.2$\,GeV and a width of $\Gamma_t=1.44262$\,GeV. We use $n_f=5$ massless quark
flavours and the corresponding NNLO PDF set with $\as(\mz{}) = 0.118$ of NNPDF3.1~\cite{Ball:2017nwa}. 
The renormalization scale of the two overall powers of the strong coupling 
is set to the Higgs mass $\mu_R^{(0)}=K_R\,m_H$. For the extra powers of the strong coupling and the factorization scale, we follow
the standard \minnlo{} scale settings described in \citere{Monni:2020nks}, which effectively sets $\mu_R\sim K_R\,\pt$ and  $\mu_F\sim K_F\,\pt$
at small transverse momentum. The central scale setting corresponds to $K_R=K_F=1$, while scale uncertainties are estimated
by varying $K_R$ and  $K_F$ up and down by a factor of two with the constraint  $1/2 \le K_R/K_F\le 2$.
For the technical settings of the \minnlo{} predictions we employ modified logarithms that smoothly turn off the resummation effects
at large transverse momenta using the setting {\tt modlog\_p 6} in the \POWHEG{} input card, with an associated scale of $Q=K_Q\,m_h$ with $K_Q=1$. And 
we  smoothly avoid the Landau singularity by multiplying the unphysical scales by a profile function at small transverse, using $Q_0=2$ \cite{Monni:2020nks}.

All \minnlo{} results are showered with \PYTHIA{8}~\cite{Sjostrand:2014zea}, switching off hadronization and underlying event.
Besides on-shell Higgs results we also study Higgs-boson decays to two photons, which we model through \PYTHIA{8} assuming a branching fraction 
of ${\rm BR}(H\to\gamma\gamma)=0.00227$. While no cuts are used in the on-shell case, for the diphoton final states we compare the fully inclusive case and 
two fiducial setups with typical selection criteria \cite{ATLAS:2022qef}. In both fiducial setups we require a rapidity threshold of $|\eta_\gamma|<1.37$ or $1.52<|\eta_\gamma|<2.37$.
For the the transverse momentum of the photons we consider both asymmetric cuts with $p_{T,\gamma_1} > 0.35\,m_H$ and $p_{T,\gamma_2} > 0.25\,m_H$, dubbed {\tt fid-asym-cuts}, and product cuts such that $\sqrt{p_{T,\gamma_1}\,p_{T,\gamma_2}} > 0.35\,m_H$ and $p_{T,\gamma_2} > 0.25\,m_H$, as suggested in \citere{Salam:2021tbm}, dubbed {\tt fid-prod-cuts}.

We obtain reference \nnlo{} results from \Matrix{} \cite{Grazzini:2017mhc} in the HTL 
with the same input setup. For the unphysical scales we set $\mu_R^{(0)}=\mu_R\sim K_R\,m_H$ and  $\mu_F\sim K_F\,m_H$, 
and the central scale and the scale variations are obtained as described above.

\begin{table}[t]
  \vspace*{0.3ex}
  \begin{center}
\begin{tabular}{l|cc}
\toprule
& $\sigma_{\rm total}$ [pb] & ratio to NNLO QCD \\ 
\midrule
NNLO QCD \;& \; $40.32(2)_{-10.7\%}^{+10.4\%}$\,& 1.000 \\
\minlo{} \;& \; $31.16(1)_{-17.6\%}^{+22.9\%}$\,& 0.773 \\
\minnlo{} \;& \; $39.55(1)_{-10.5\%}^{+11.0\%}$\,& 0.981 \\
\bottomrule
\end{tabular}
\end{center}
  \caption{
    Total cross section in the HTL at NNLO QCD compared to \minlo{} and \minnlo{}.\label{tab:HTLXS}}
\end{table}

\begin{figure}[t]
\begin{center}
\begin{tabular}{cc}\vspace*{-0.3cm}
\includegraphics[width=.46\textwidth]{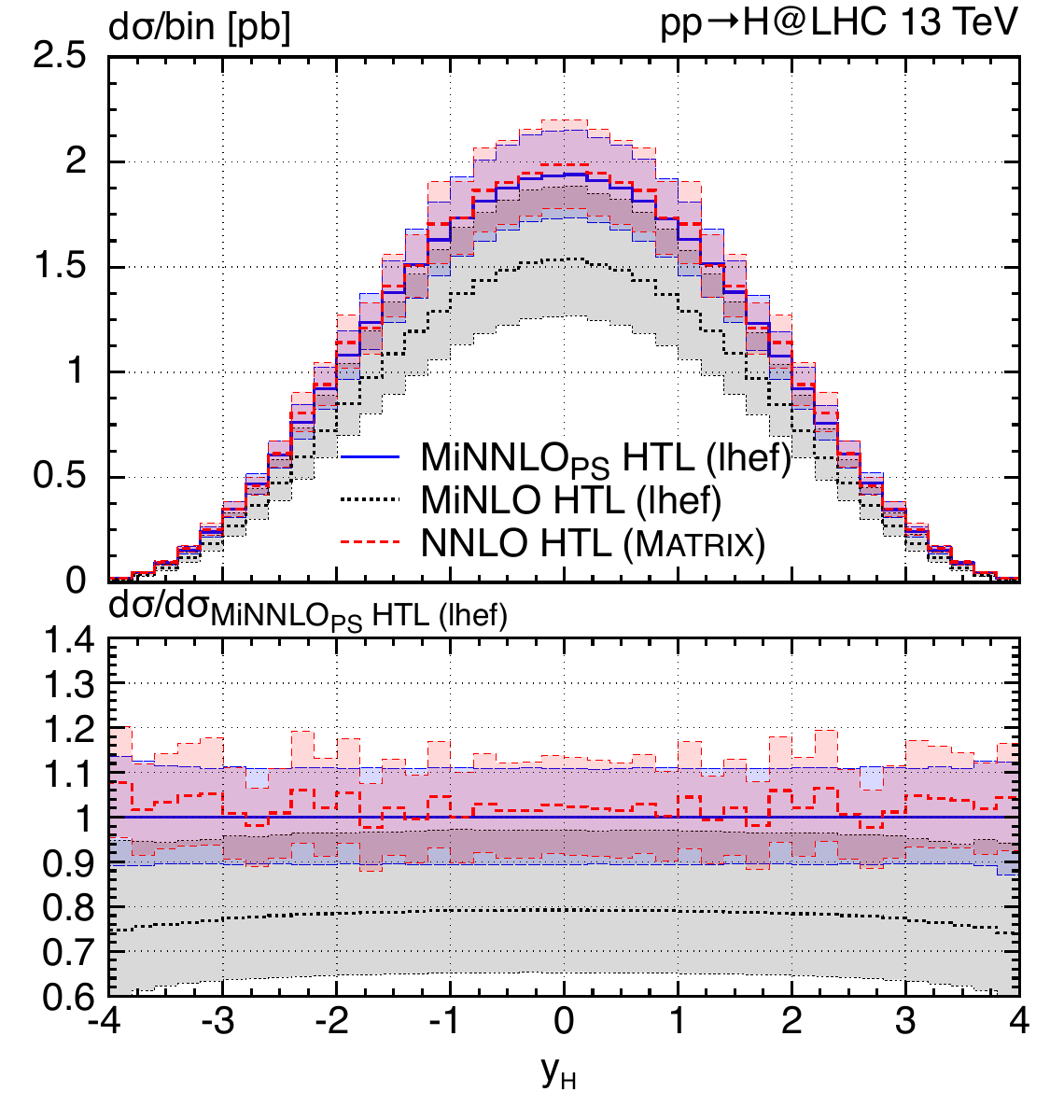}
&
\includegraphics[width=.46\textwidth]{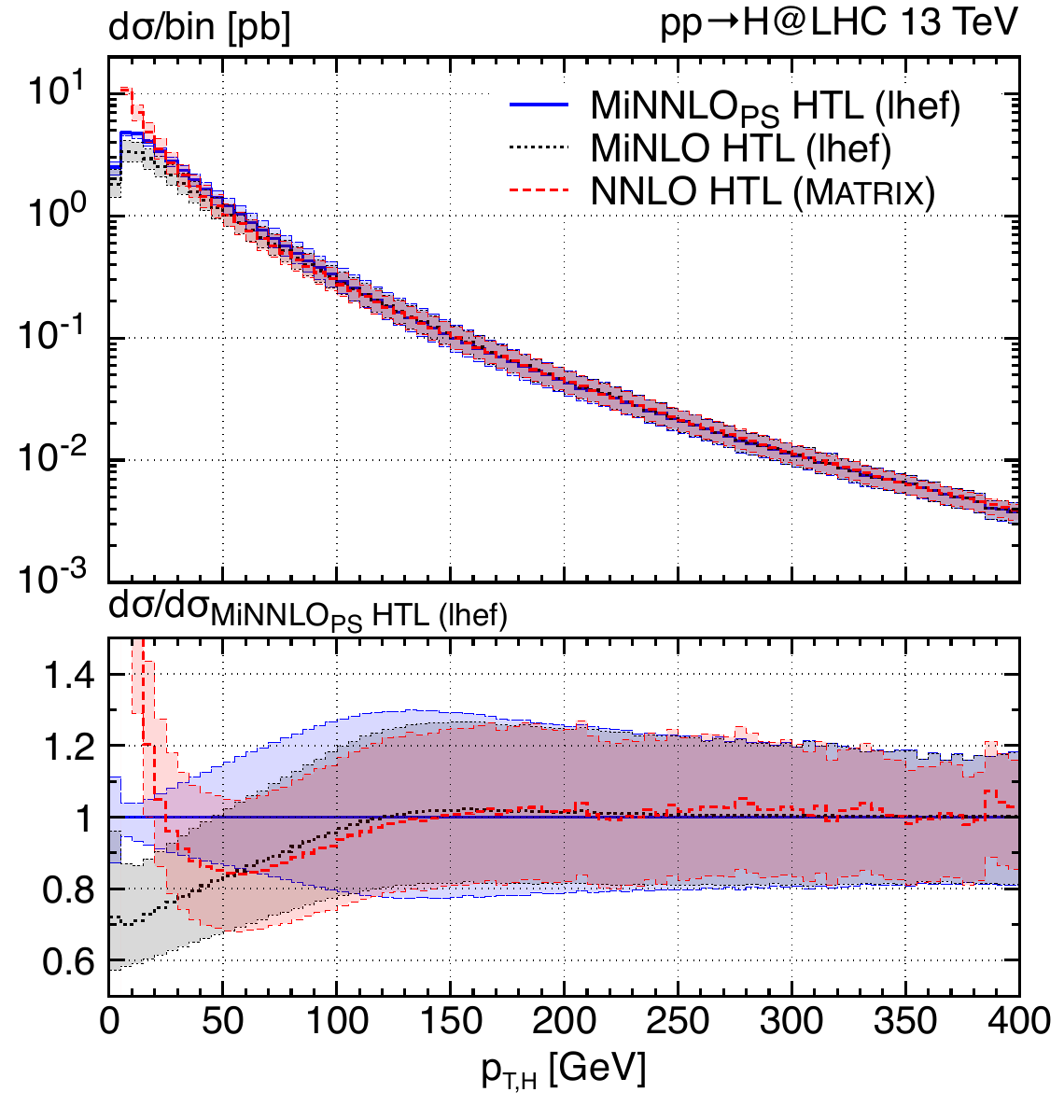}
\end{tabular}
\vspace*{1ex}
\caption{\label{fig:validation} Comparison of \minnlo{} and \minlo{} predictions with fixed-order NNLO results.}
\end{center}
\end{figure}

We start by validating our new implementation of the \minnlo{} $H$ generator in \POWHEGBOXRES{} 
against fixed-order NNLO QCD predictions in the HTL. Shower effects are turned off 
in this comparison by using results at Les-Houches-Event (LHE) level for a more direct comparison to fixed order.
The total cross sections at fixed-order NNLO QCD and predicted by \minnlo{}, given in \tab{tab:HTLXS},
lie within 2\%, and therefore agree well within the given uncertainties. We recall that the two predictions differ 
in the treatment of terms beyond accuracy and are not expected to yield the same numerical result. 
The \minlo{} cross section, on the other hand, is more than 20\% lower, and features substantially larger 
scale uncertainties.

In \fig{fig:validation}, we show the same comparison for differential distributions.
The NNLO QCD prediction of \Matrix{} is
represented by a red, dashed curve with a red band, while the \minlo{} and
\minnlo{} predictions are shown in black, dotted and blue, solid, respectively.
For the rapidity distribution of the Higgs boson ($y_H$) in the left figure, we observe 
excellent agreement within the quoted scale uncertainties between the \minnlo{} and NNLO QCD histograms.
The scale uncertainties are of the same size (about $\sim 10\%$) and the central values agree at the level 
of 1-2\%. Compared to the \minlo{} prediction we observe relatively flat \minnlo{} corrections of about $+25\%$ 
and a smaller scale-uncertainty band. For the Higgs transverse-momentum ($p_{T,H}$) spectrum in the right figure,
on the other hand, \minnlo{} and NNLO QCD results agree at large $p_{T,H}$. At small $p_{T,H}$ the fixed-order 
prediction diverges due to the large logarithmic contributions, while \minlo{} and \minnlo{} predictions remain finite.
Note that at large $p_{T,H}$ all three predictions are formally only NLO accurate and coincide with each other.
It is interesting to observe the typical pattern that the corrections included through \minnlo{} with respect to \minlo{}
all enter the cross section at small $p_{T,H}$ and are then smoothly turned off. 
This is physically appropriate and a direct consequence of the NNLO+PS matching approach and the employed modified logarithms.

\begin{table}[t]
  \vspace*{0.3ex}
  \begin{center}
\begin{tabular}{l|cc}
\toprule
& $\sigma_{\rm total}$ [pb] & ratio to HTL \\ 
\midrule
HTL \;& \; $39.55(1)_{-10.5\%}^{+11.0\%}$\,& 1.000 \\
HEFT \;& \; $42.13(1)_{-10.5\%}^{+11.0\%}$\,& 1.065 \\ 
FT \;& \; $42.01(1)_{-10.6\%}^{+11.2\%}$\,& 1.062 \\
 FT-approx-2  \;& \; $41.73(1)_{-10.4\%}^{+10.9\%}$\, & 1.055 \\ 
FT-approx-3  \;& \; $41.76(1)_{-10.4\%}^{+10.9\%}$\, & 1.056 \\ 
\bottomrule
\end{tabular}
\end{center}
  \caption{Total cross sections predicted by \minnlo{} in various approximations. \label{tab:XS}}
\end{table} 

\begin{figure}[t!]
\begin{center}
\begin{tabular}{cc}\vspace*{-0.3cm}
\includegraphics[width=.46\textwidth]{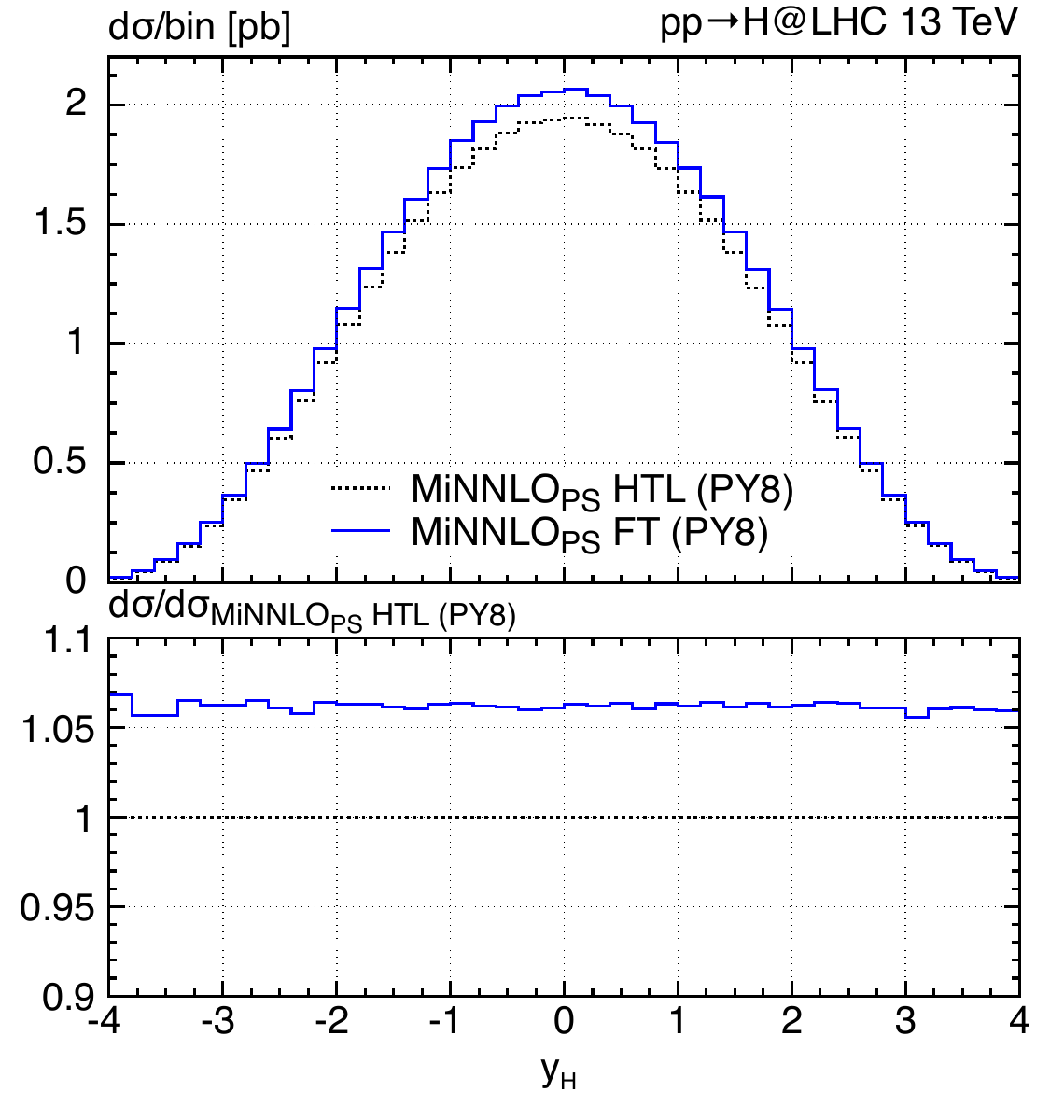}
&
\includegraphics[width=.46\textwidth]{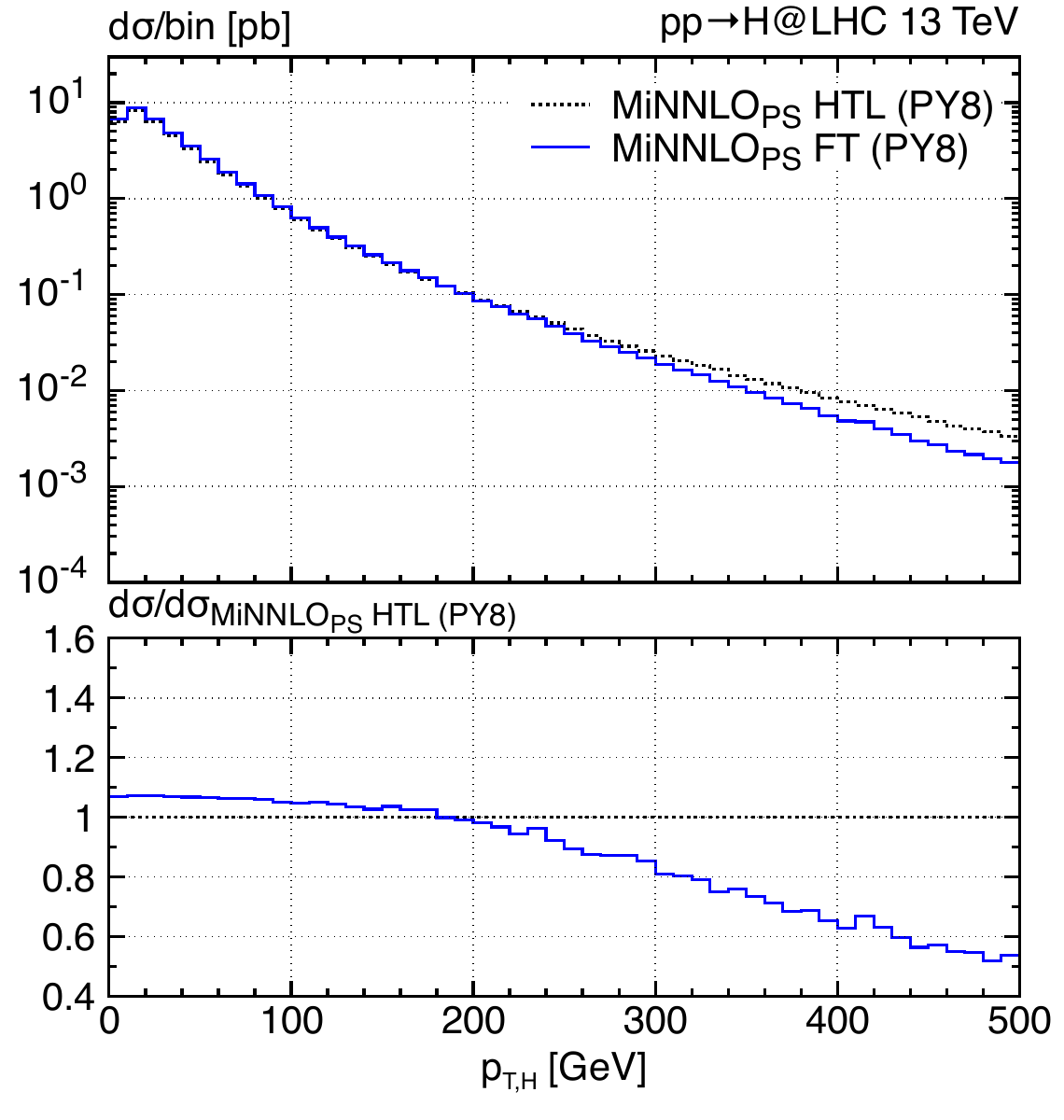}\\
\includegraphics[width=.46\textwidth]{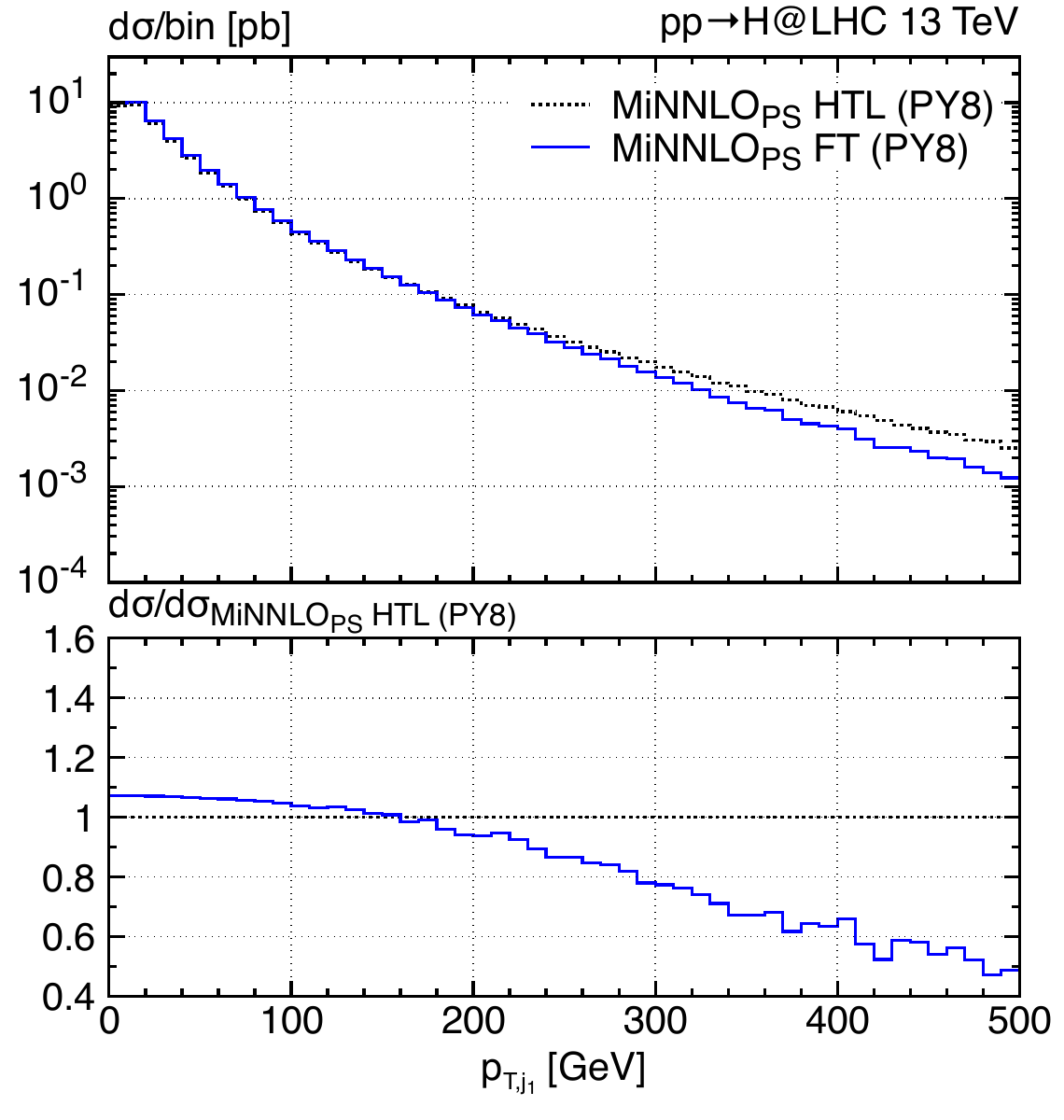}
&
\includegraphics[width=.46\textwidth]{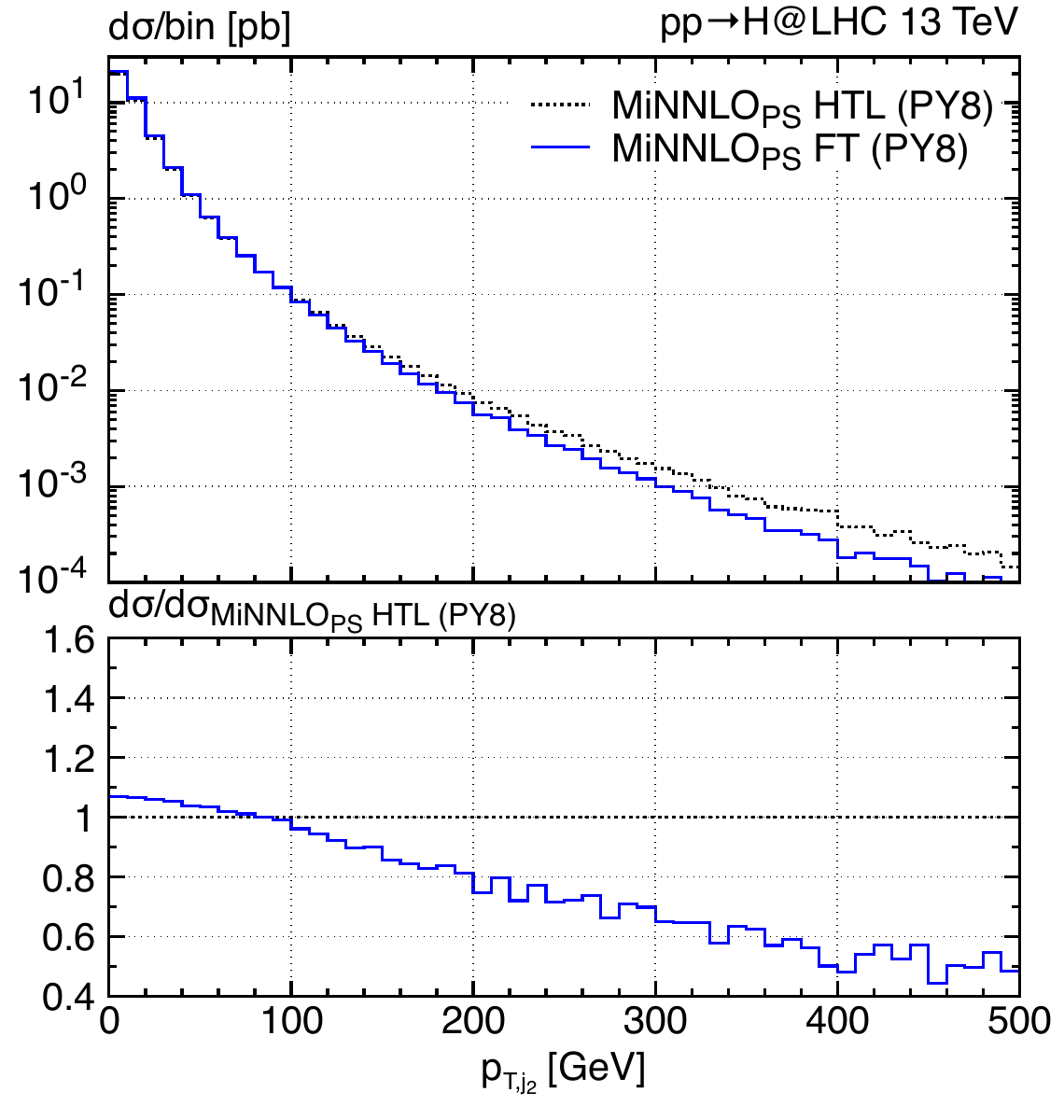}

\end{tabular}
\vspace*{1ex}
\caption{\label{fig:topmass} Top-mass effects in the FT compared to the HTL for \minnlo{} predictions.}
\end{center}
\end{figure}

After having validated our  \minnlo{} event generator for Higgs-boson production, we now analyze the effect of including the
exact top-mass dependence at NNLO in QCD. \tab{tab:XS} shows the total inclusive cross section predicted by our \minnlo{} generator
in various approximations, HTL, HEFT, FT, FT-approx-2, and FT-approx-3. The HEFT result corresponds to the HTL result rescaled
by exact top-mass dependence at LO, as given in \eqn{eq:HEFT}. Comparing the full theory (FT) result against the HTL one, we find mass effects
of about 6.2\%. Accounting for the mass effects at LO, and thus comparing the FT prediction against the HEFT approximation, we see that the
FT is lower than the HEFT one by only -0.3\%. This confirms the effect that has been found in the study of the total inclusive
NNLO QCD cross section at fixed order in \citere{Czakon:2021yub}. Moreover, we can also conclude from \tab{tab:XS}, that the approximations
introduced in \sct{sec:approx}, namely FT-approx-2 and FT-approx-3, provide a good approximation of the top-mass effects for the total inclusive 
cross section, being only 0.6--0.7\% lower than the FT cross section.\footnote{Notice that FT-approx-1 is so close to the other two results that we refrain from including it here and in the following.}

We continue by analyzing the mass effects in differential distributions  in \fig{fig:topmass}. Note that the full theory (FT) results, shown as a blue, solid curve, 
have been obtained from the HTL results, shown in black, dotted curve, by reweighting the Les-Houches events to the complete calculation, 
since the FT matrix elements are substantially slower. This approach is exact in the limit of large statistics, bearing that
it does not entail any approximation beyond the (very small) numerical uncertainties.
The top-quark mass effects can be read off the blue FT/HTL ratio curve in the lower panels of the figures. In the $y_H$ distribution 
the effect is completely flat and about $+6\%$, as already observed for the total inclusive cross section in \tab{tab:XS}.
In the transverse momentum distribution of the Higgs boson (upper right figure), of the leading jet ($p_{T,j_1}$, lower left figure)
and of the subleading jet ($p_{T,j_2}$, lower right figure), the top-mass effects show a very similar behaviour. At small $p_T$ the effect is about $+6\%$, while 
the top-mass dependence steadily decreases the cross section in the transverse-momentum tails, halving the \minnlo{} cross section in the 
HTL approximation around $500$\,GeV. This feature is well known and related to the fact that at large transverse momentum the top quarks in 
the loop get resolved, leading to a smaller cross section.

\begin{figure}[t]
\begin{center}
\begin{tabular}{cc}\vspace*{-0.3cm}
\includegraphics[width=.46\textwidth]{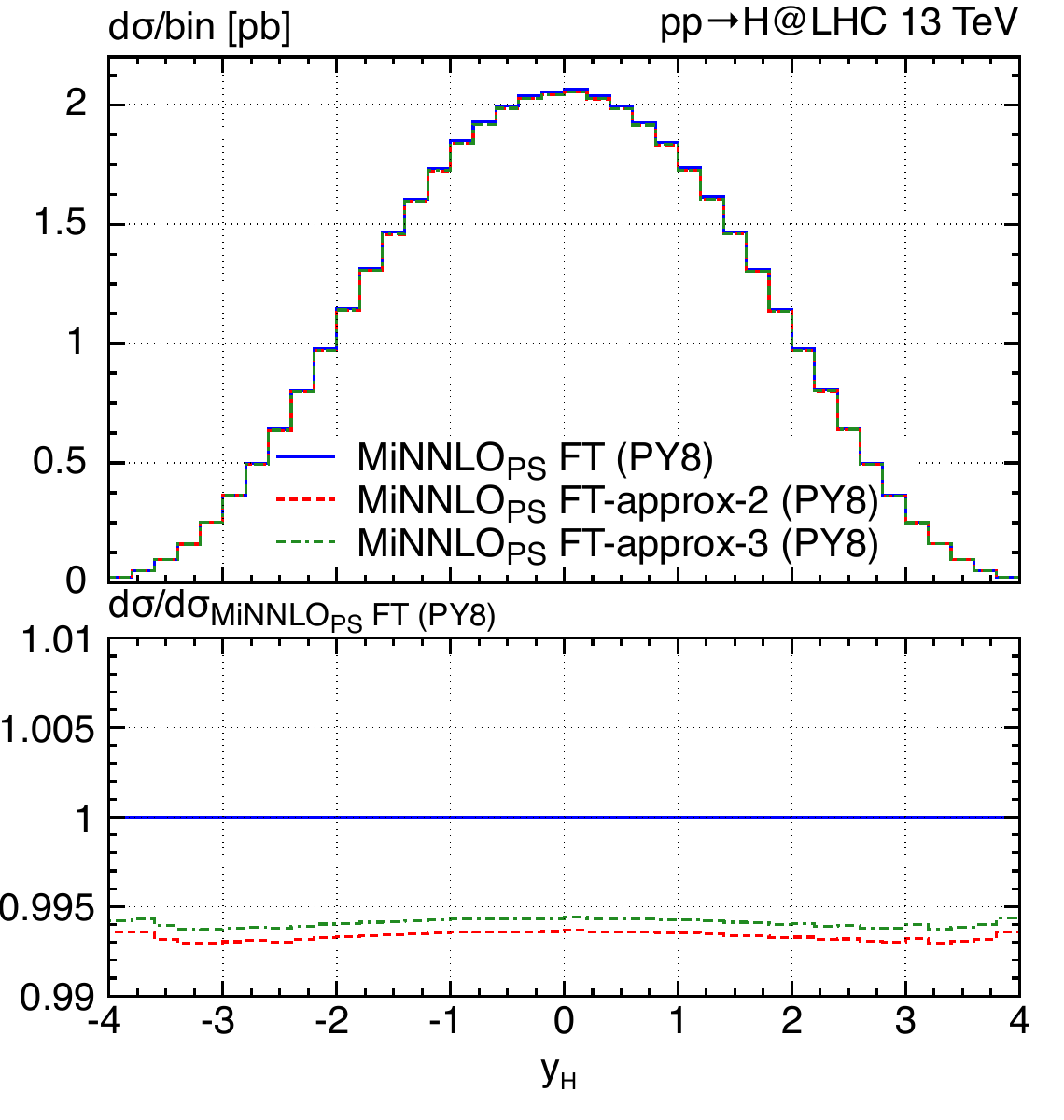}
&
\includegraphics[width=.46\textwidth]{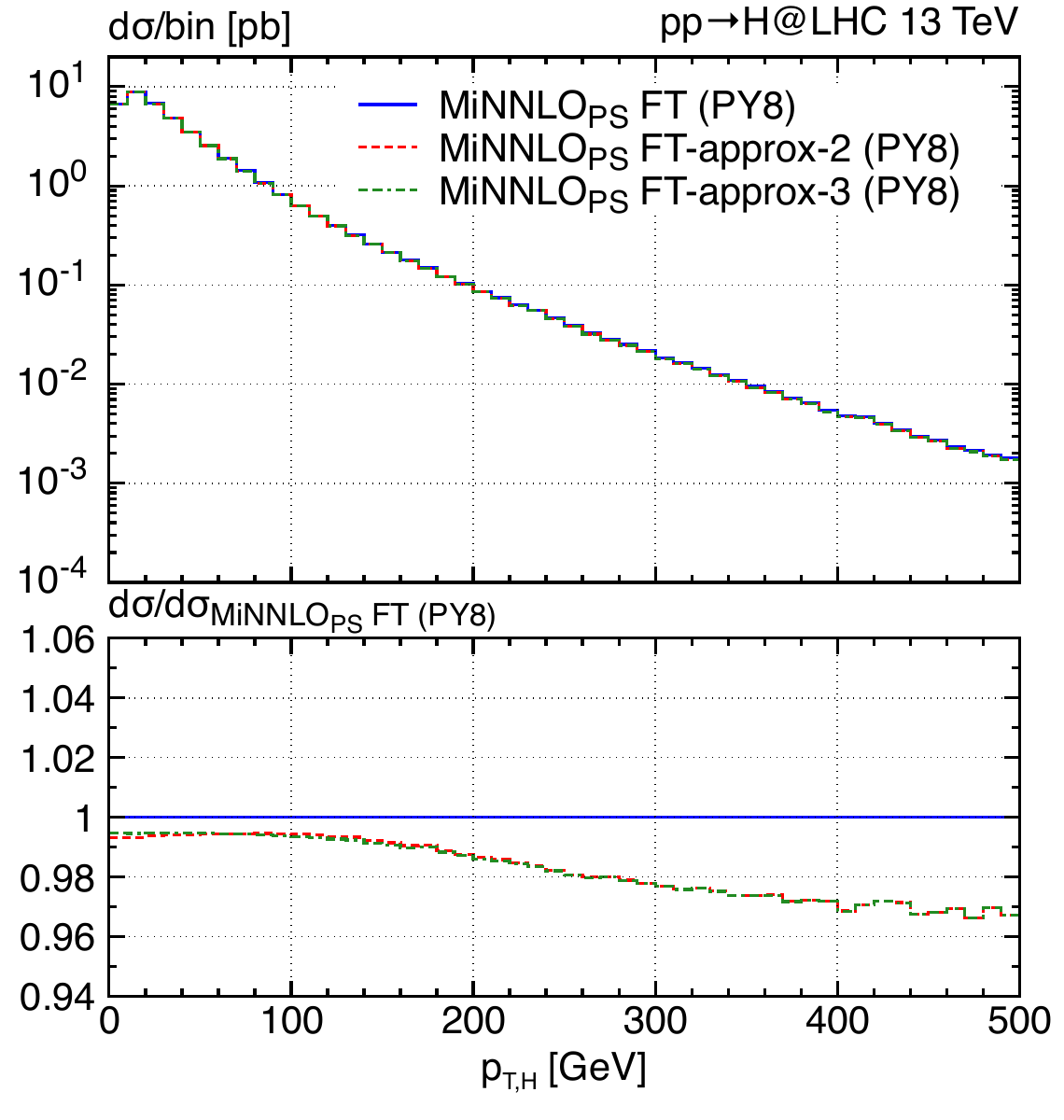}
\end{tabular}
\vspace*{1ex}
\caption{\label{fig:FTapprox} Comparison of top-mass approximations to the FT result for \minnlo{} predictions.}
\end{center}
\end{figure}

Next, we would like to assess the quality of the different approximations of the top-mass dependence in the \minnlo{} cross section, defined in \sct{sec:approx},
which could be applied already before our calculation in the full theory of this manuscript. To this end, we compare the complete top-mass dependence in
\minnlo{} (blue, solid) with the corresponding implementations of the approximations FT-approx-2 (red, dashed curve) and FT-approx-3 (green, double-dash-dotted curve) 
in \fig{fig:FTapprox}. We find that for inclusive NNLO observables, i.e. the inclusive 
cross section and the Higgs rapidity distribution (left figure), the approximations of the FT work perfectly and differ from the full result by only about six permille.
This also holds true for small $p_{T,H}$ values. On the other hand, we can observe that at large $p_{T,H}$ the approximations successively differ more from the exact result,
reaching about 3\% at $p_{T,H}=500$\,GeV. Nevertheless, the approximations work at a very satisfactory level, especially considering the scale uncertainties, which are much larger.
The corresponding figures for other transverse-momentum observables, like $p_{T,j_1}$ and $p_{T,j_2}$, show the same (relative) behaviour as for $p_{T,H}$, which is why 
we refrain from showing them here, and, thus, the same conclusions apply to them.

\begin{figure}[t]
\begin{center}
\begin{tabular}{ccc}\vspace*{-0.3cm}
\includegraphics[width=.31\textwidth]{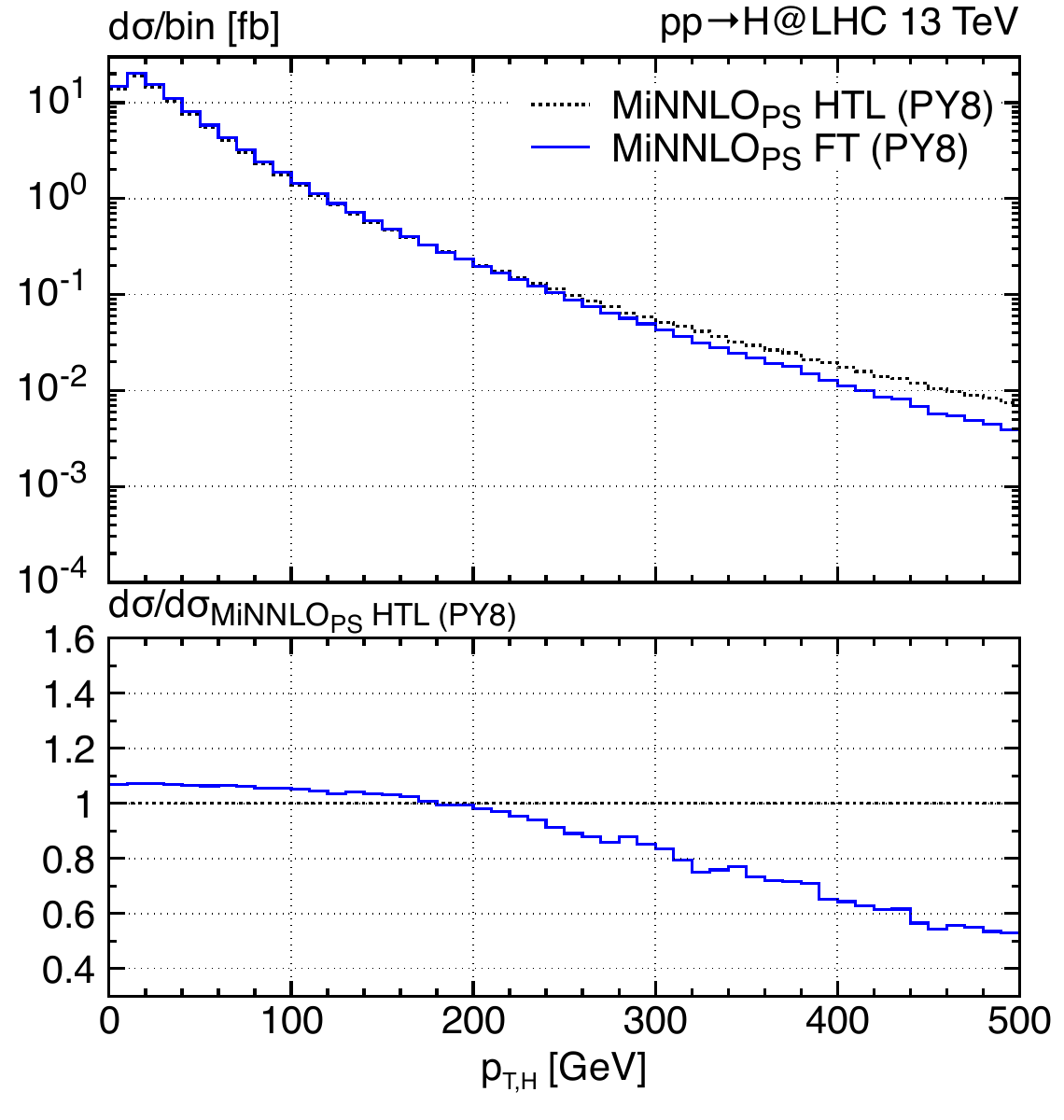}
&
\includegraphics[width=.31\textwidth]{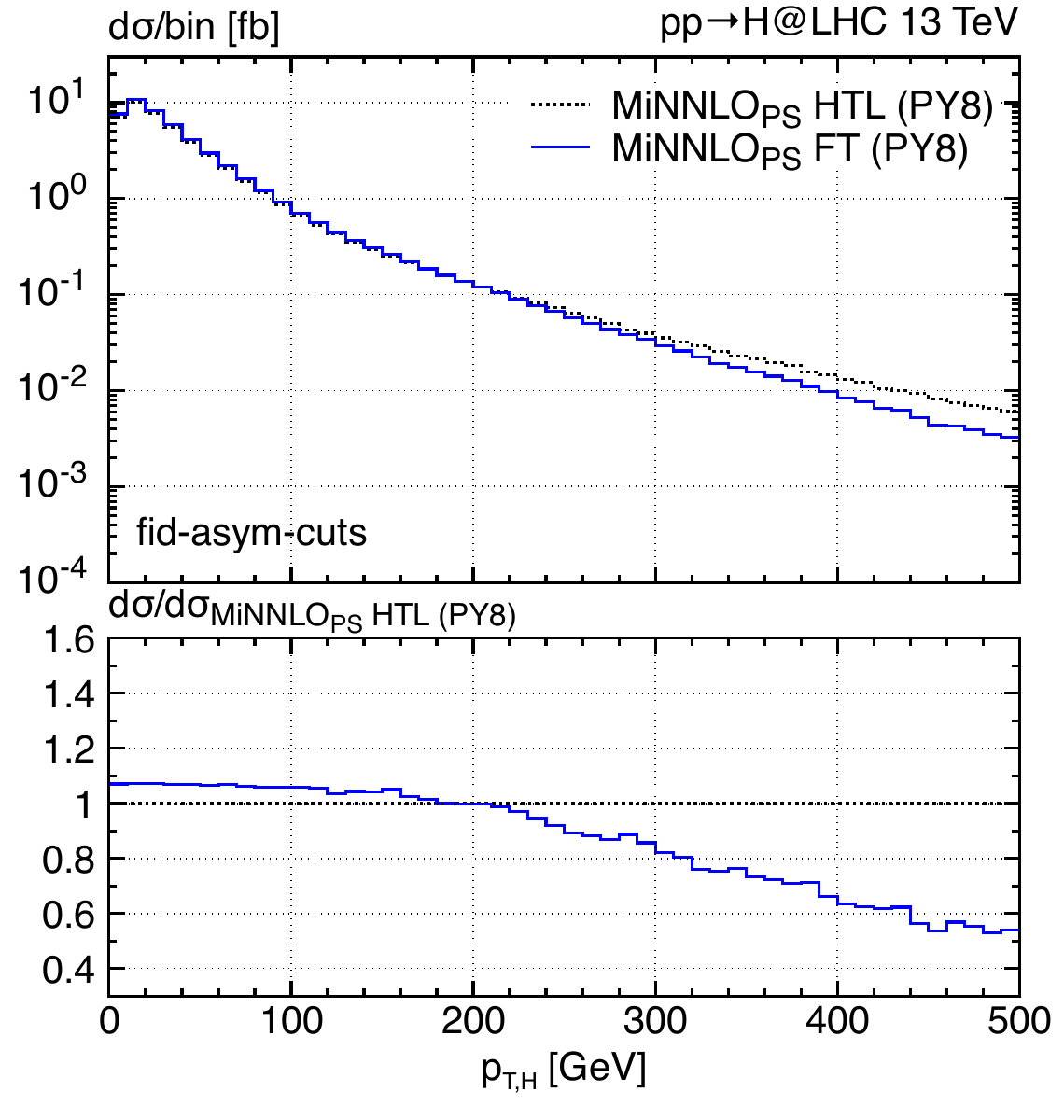}
&
\includegraphics[width=.31\textwidth]{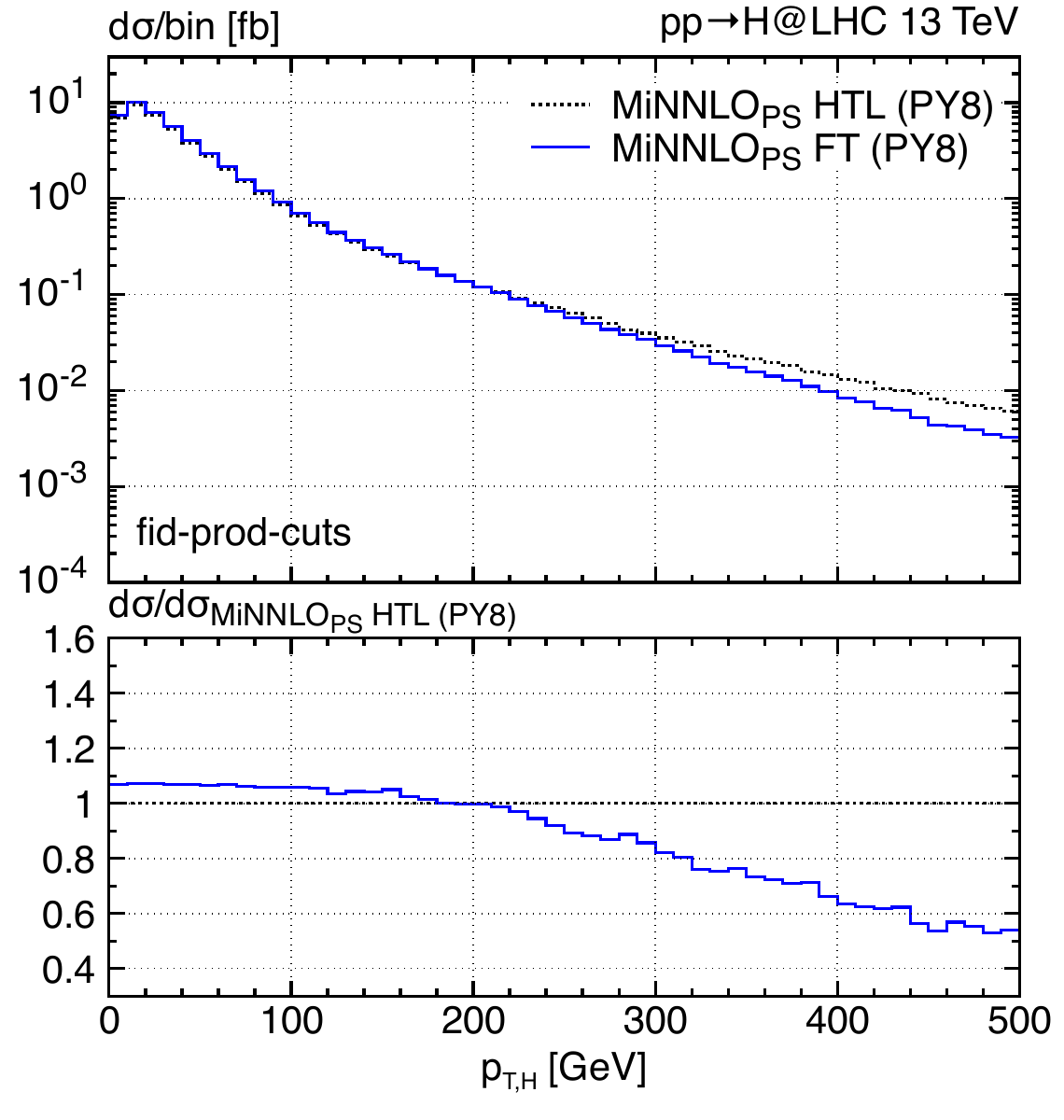}\\
\includegraphics[width=.31\textwidth]{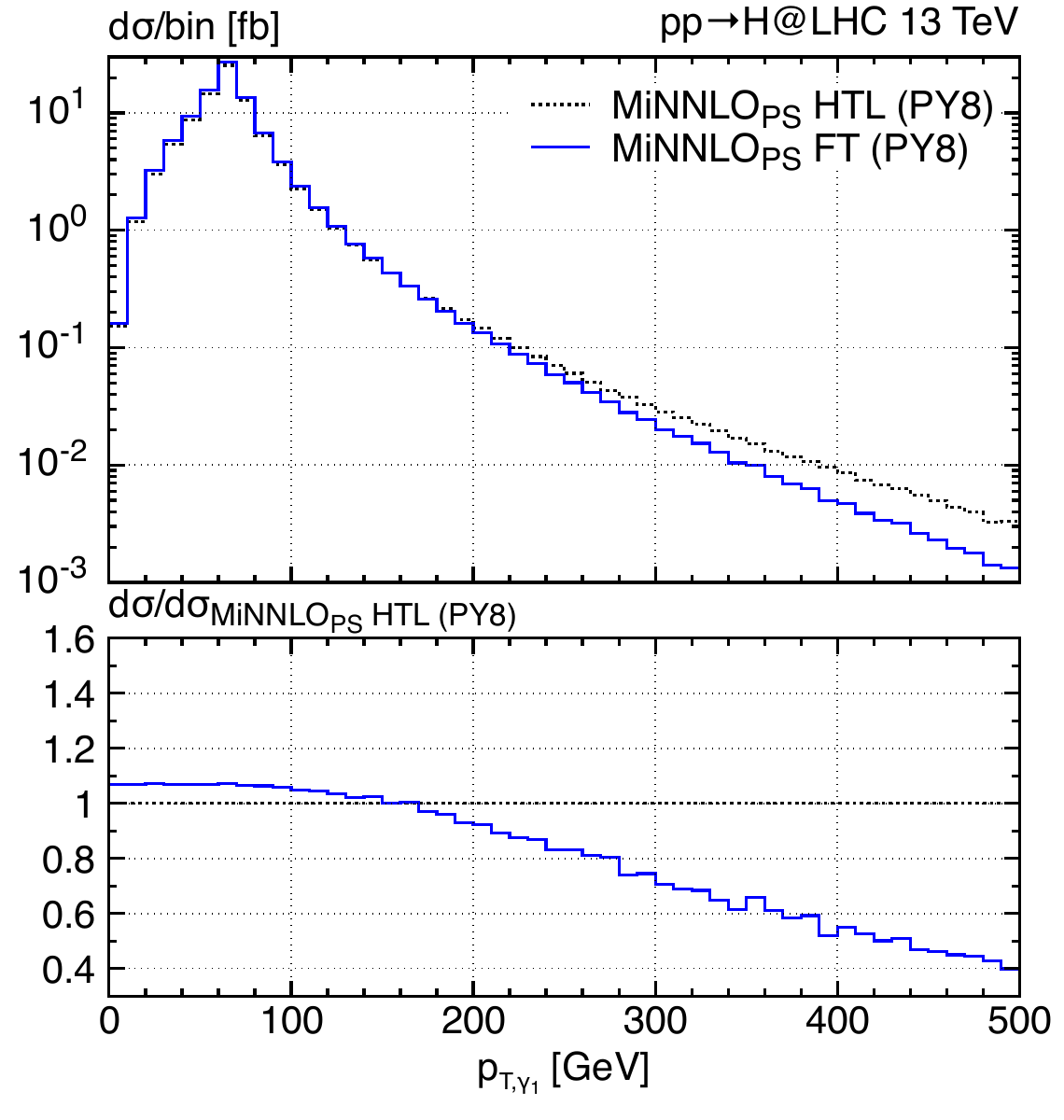}
&
\includegraphics[width=.31\textwidth]{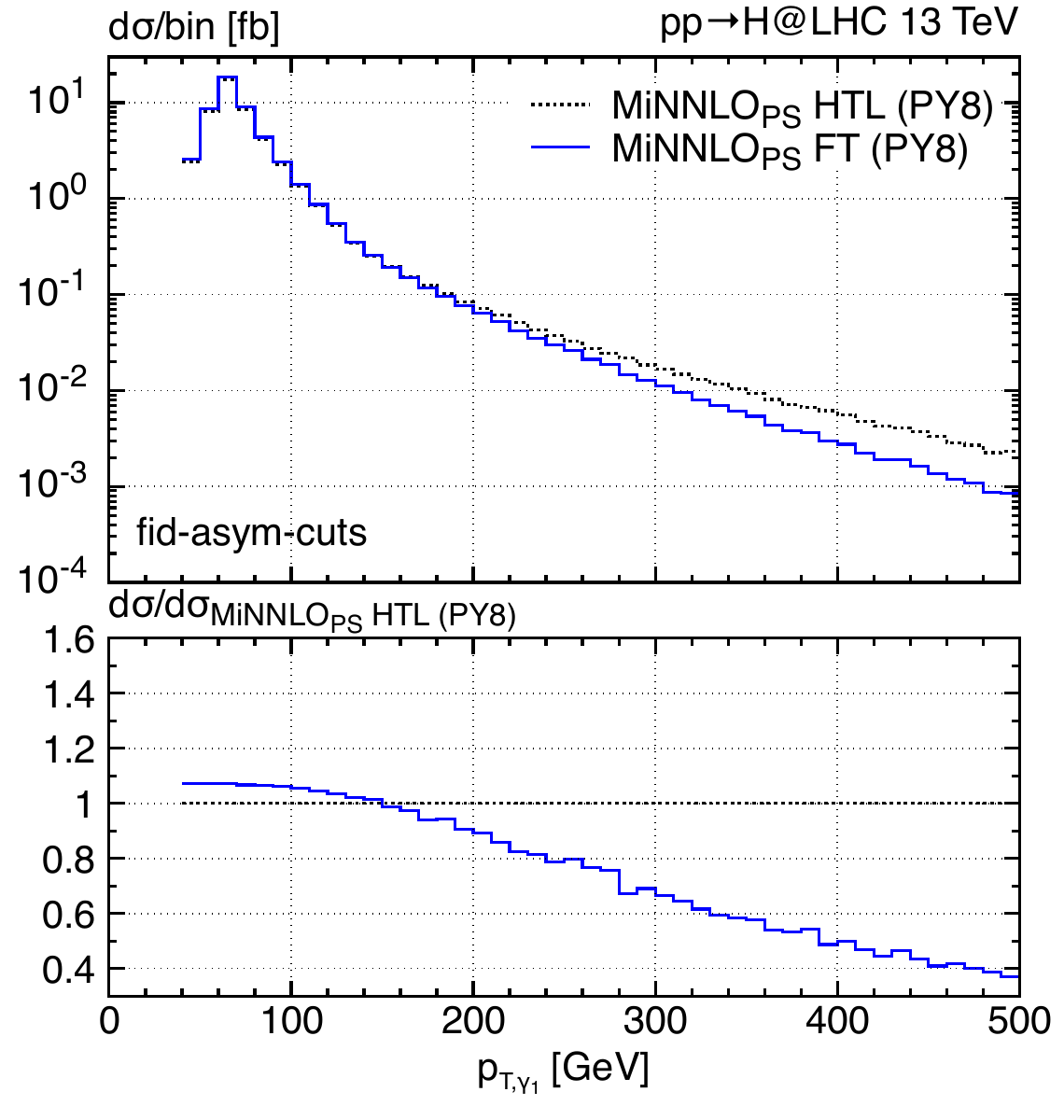}
&
\includegraphics[width=.31\textwidth]{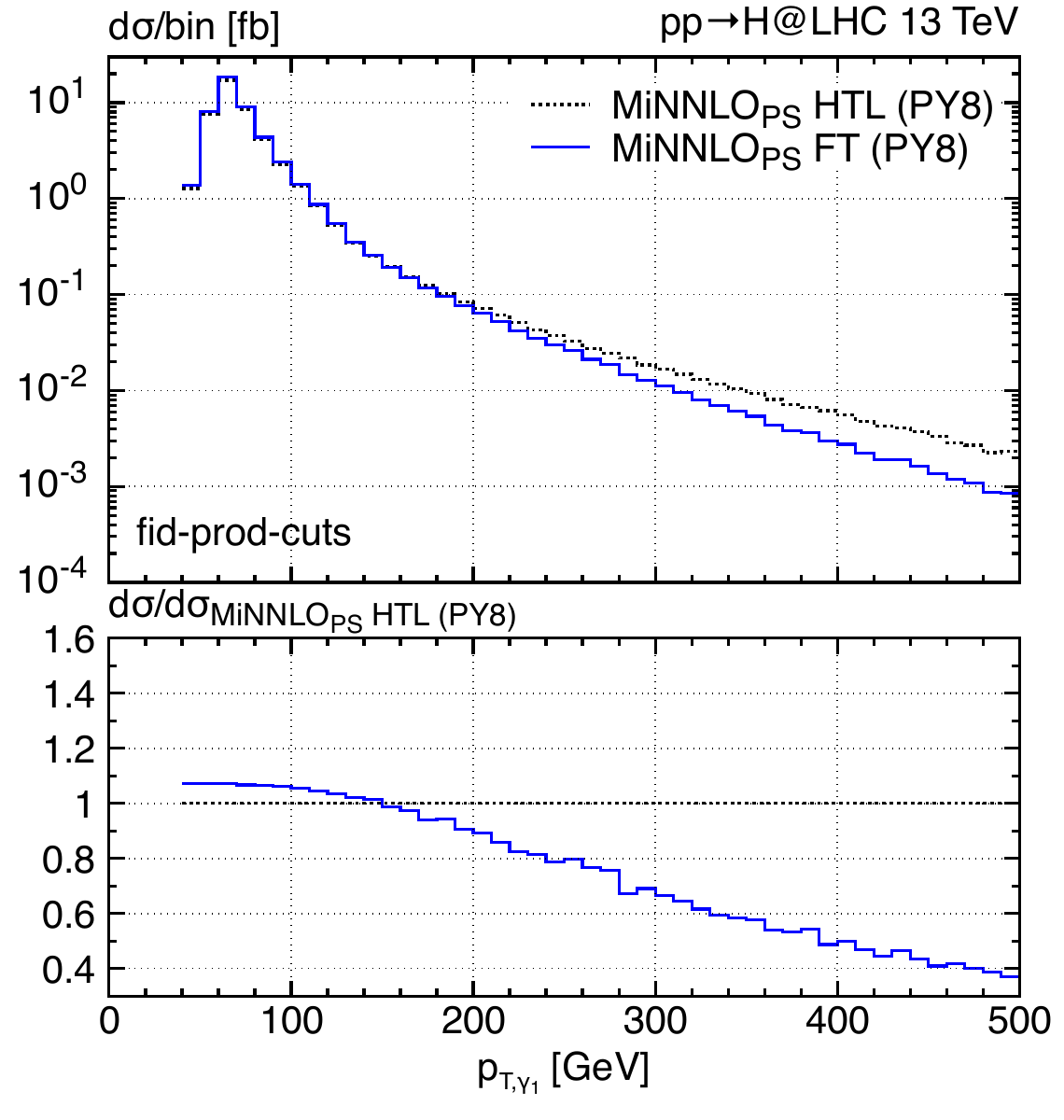}
\end{tabular}
\vspace*{1ex}
\caption{\label{fig:masseffectsdiphoton} Top-mass effects in distributions of diphoton predictions obtained via \minnlo{}.}
\end{center}
\end{figure}

We now move on from discussing results for stable Higgs bosons to including the Higgs decay to photons.
In \fig{fig:masseffectsdiphoton} we consider including the full top-mass dependence compared to the HTL
by showing representative results for the $p_{T,H}$ spectrum, where the Higgs
boson is reconstructed from the two photons, and for the transverse-momentum distribution of the leading 
photon ($p_{T,\gamma_1}$). Very similar top-mass effects can be observed for other transverse-momentum 
observables of the jets and the subleading photon. Three different scenarios are considered for applying 
fiducial cuts on the photons, as introduced above: no cuts (left column), with asymmetric cuts on the photons ({\tt fid-asym-cuts}, center column),
with product cuts on the photons ({\tt fid-prod-cuts}, right column). 
We observe that the size of top-mass effects does not depend on the cuts applied to the photons in these scenarios. 
Moreover, we find a slightly larger dependence on the top mass in the tail of the $p_{T,\gamma_1}$ distribution compared to the $p_{T,H}$ spectrum,
but the qualitative behaviour is very similar.

\begin{figure}[t]
\begin{center}
\begin{tabular}{ccc}\vspace*{-0.3cm}
\includegraphics[width=.31\textwidth]{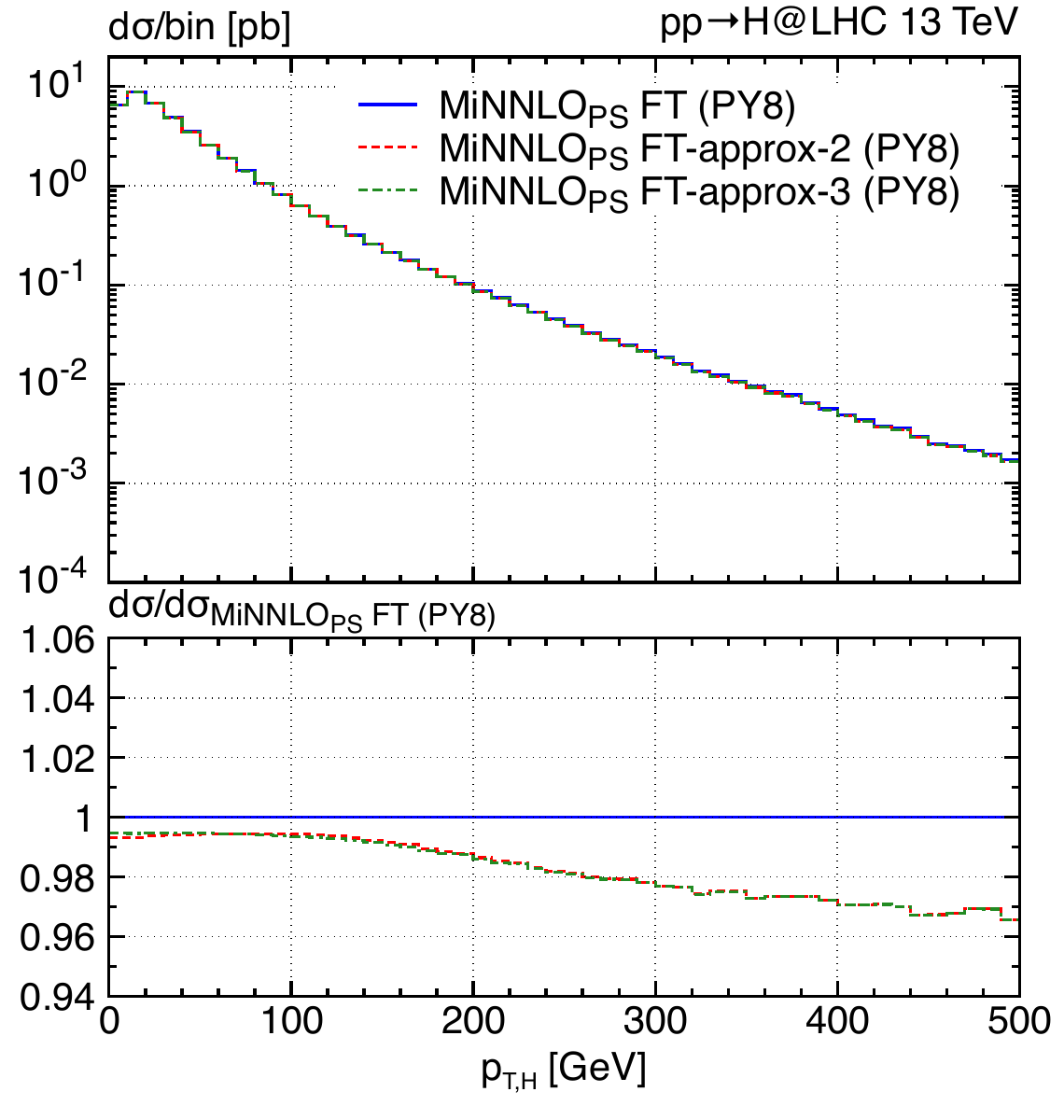}
&
\includegraphics[width=.31\textwidth]{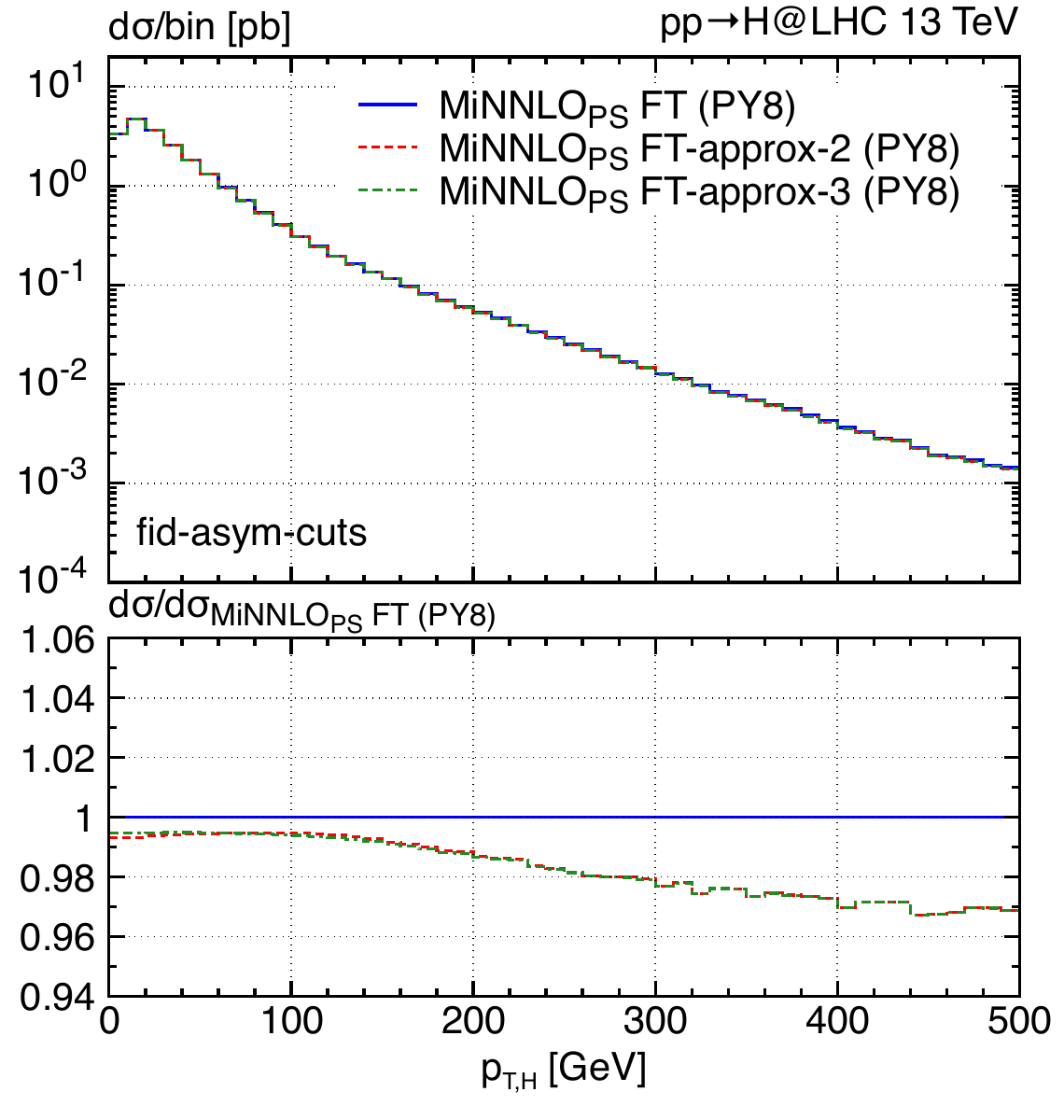}
&
\includegraphics[width=.31\textwidth]{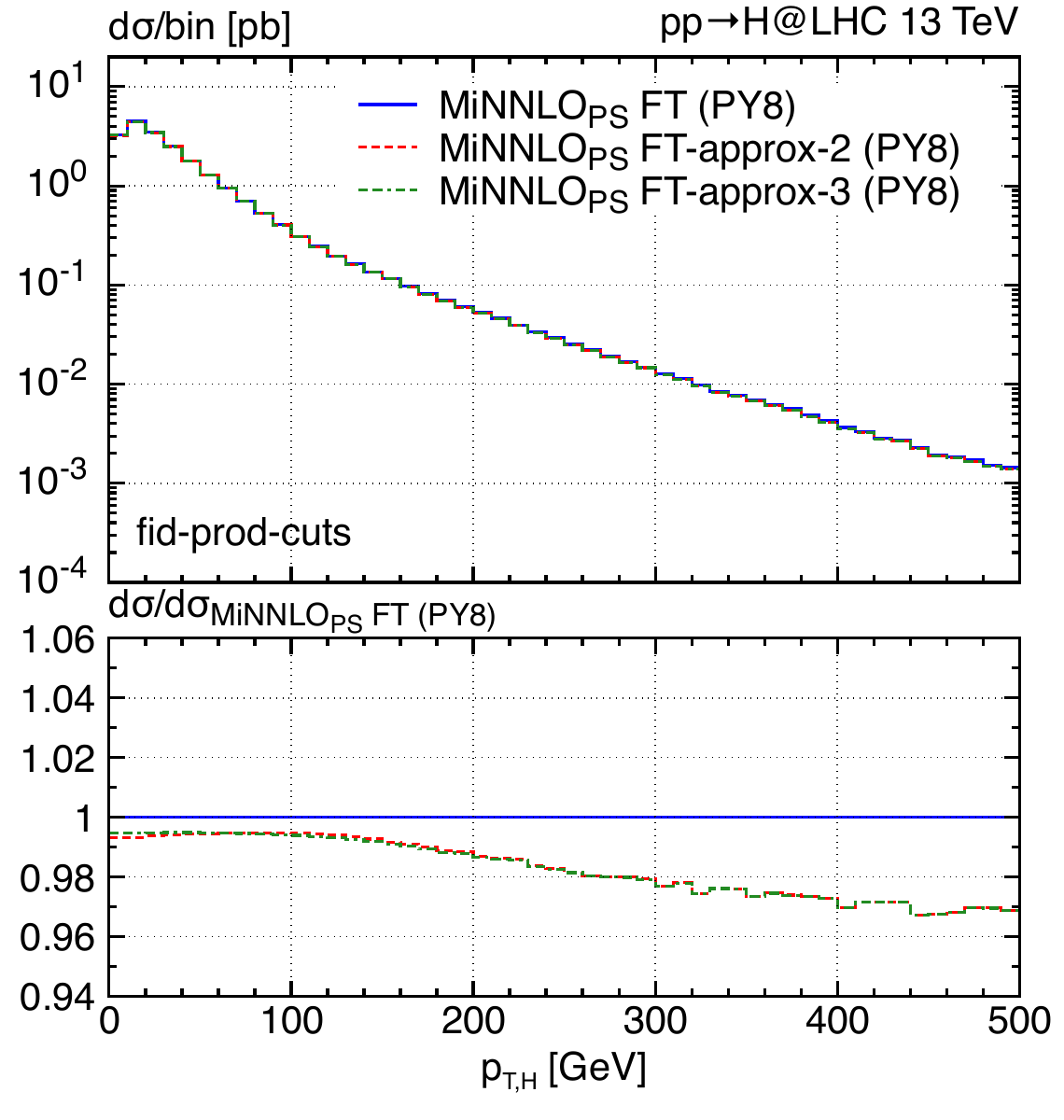}\\
\includegraphics[width=.31\textwidth]{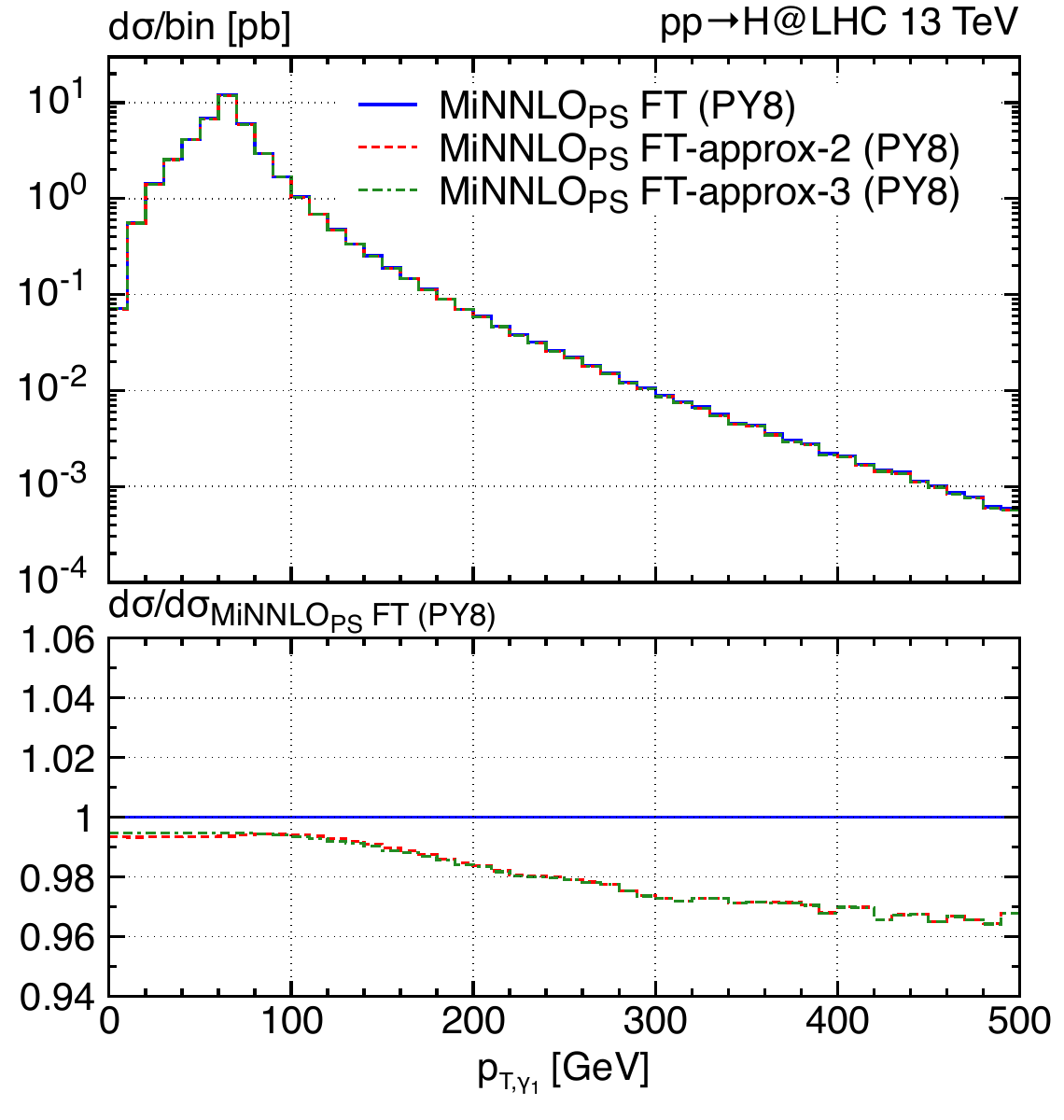}
&
\includegraphics[width=.31\textwidth]{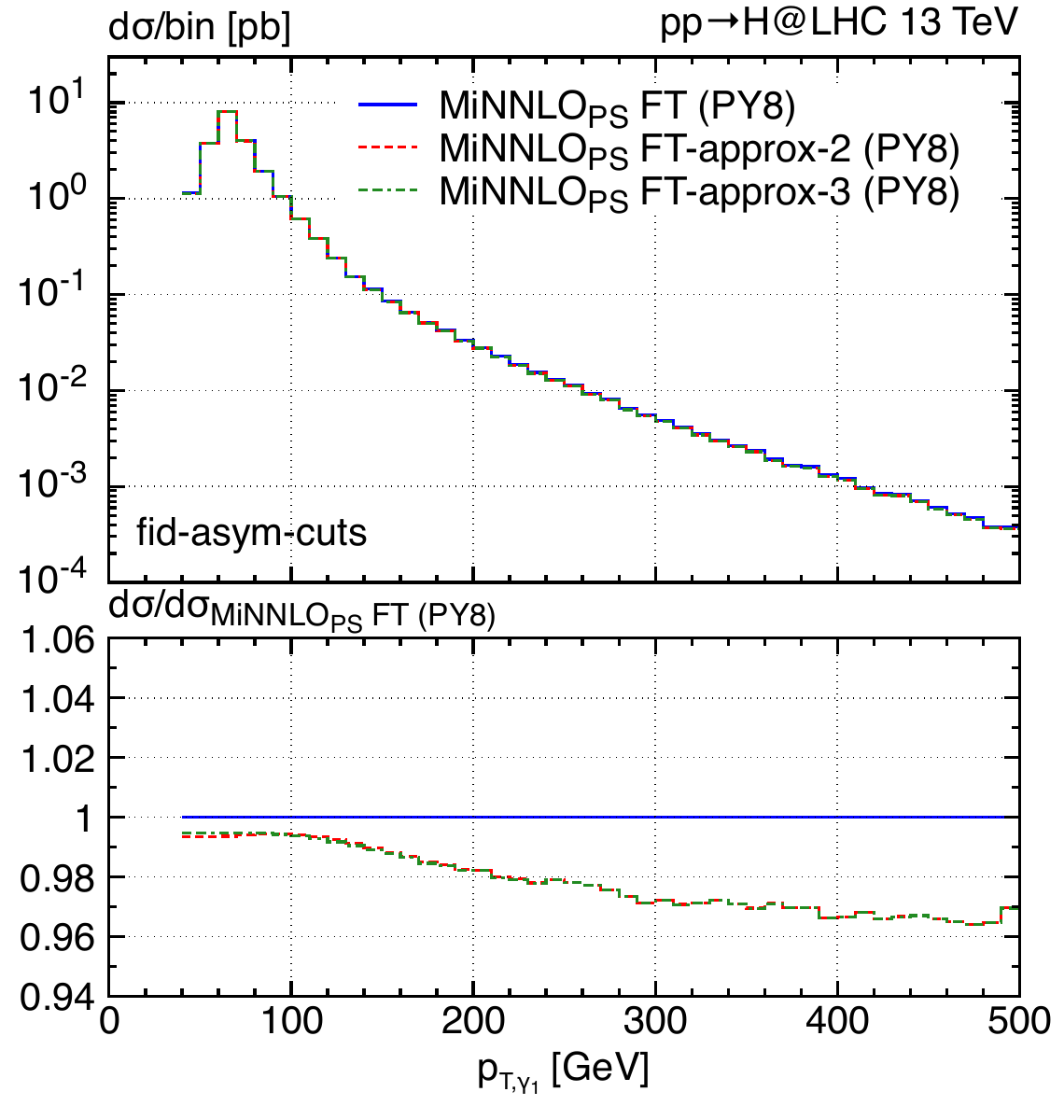}
&
\includegraphics[width=.31\textwidth]{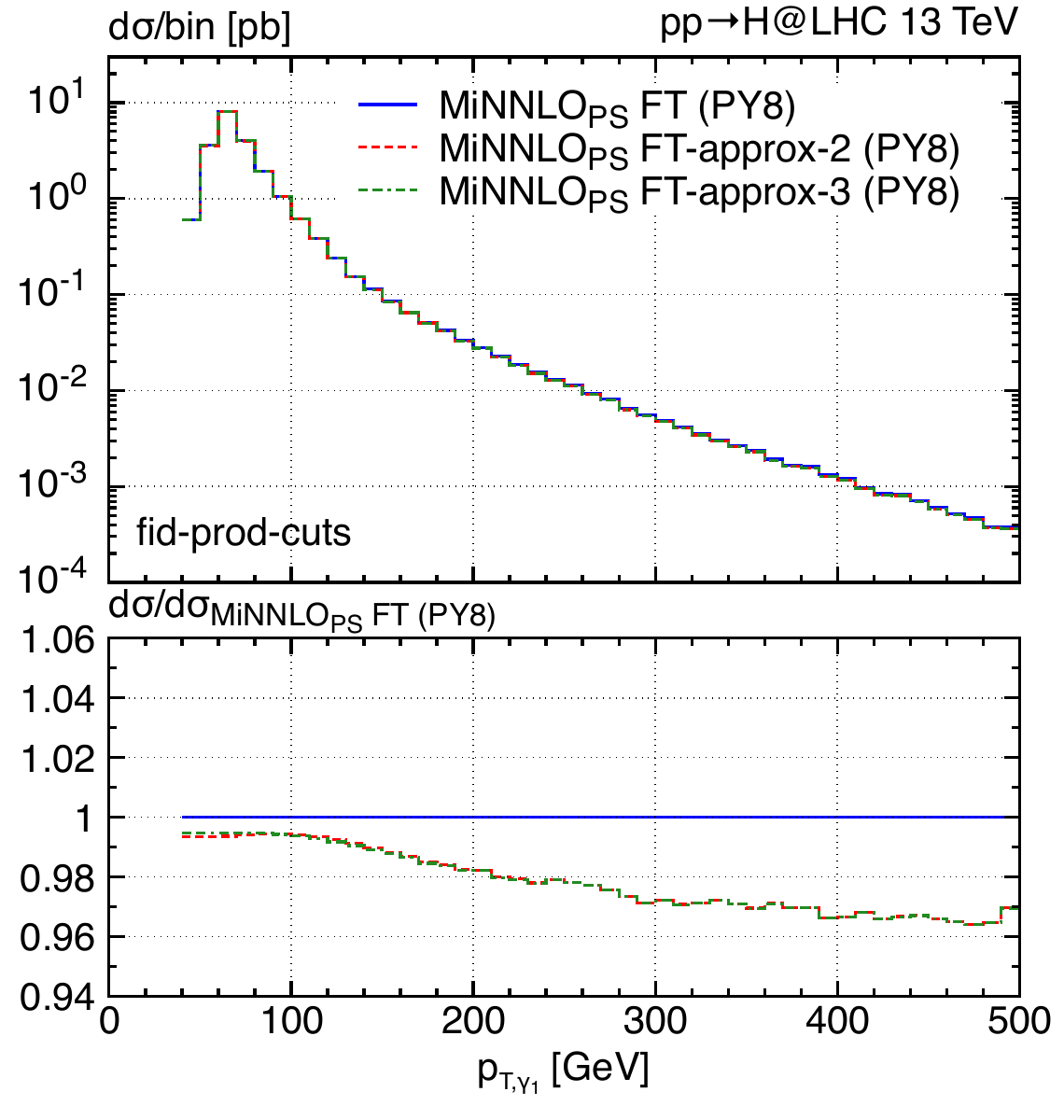}
\end{tabular}
\vspace*{1ex}
\caption{\label{fig:approxdiphoton} Quality of top-mass approximations in diphoton distributions of \minnlo{} predictions.}
\end{center}
\end{figure}

Finally, we show in \fig{fig:approxdiphoton} that also the quality of the different approximations of the top-mass effects does not 
depend on the fiducial selection cuts applied to the photons. Moreover, the approximations of the FT result work slightly worse
for the $p_{T,\gamma_1}$ spectrum compared to the $p_{T,H}$ one. The difference to the FT prediction reaches about 4\%
for  $p_{T,\gamma_1}=500$\,GeV, while it is about 3\% for $p_{T,H}=500$\,GeV.

\section{Summary}
\label{sec:summary}

We have presented the calculation of Higgs-boson production through gluon fusion in the full theory
at NNLO in QCD and matched it to a parton shower with the \minnlo{} method. For the first time,
NNLO QCD corrections are computed at the fully differential level without any approximations
for the top-quark loop, i.e.\ taking into account  the complete dependence on the top-quark mass. 
Hence, our calculation includes $gg\to H$ amplitudes up to three loops and 
$pp\to H$+jet amplitudes up to two loops in the full SM theory.

We have studied the effect of the top-quark mass  on our \minnlo{} predictions with respect to 
the approximation of an infinitely heavy top quark both for a stable Higgs boson and including its decay to photons. 
We find the impact of the top-quark mass to be about +6\% on the total inclusive cross section and completely flat in the Higgs rapidity spectrum at NNLO QCD.
Since the relative impact is the exact same as the one already at the leading order, a rescaling of the HTL with Born 
$gg\to H$ amplitude provides an excellent approximation of the full theory result for NNLO observables.
Considering transverse-momentum spectra of the Higgs boson, the jets or the decay photons, on the other hand, 
the top-quark mass yields a large negative effect that continuously increases in the tail of these distributions,
reducing the cross section by a factor of two and more at transverse-momentum scales of around 500\,GeV.
We have also shown that these large effects in the full SM theory can be largely captured through various approximations 
that employ a rescaling of the HTL amplitudes with lower-order amplitudes including the complete top-mass dependence.
The differences between these approximations
and the full result are less than 5\% around transverse momenta of 500\,GeV and only at the few-permille level for inclusive
NNLO observables, which is very small compared to the uncertainties due to missing higher-order corrections.

Our analysis reflects that it is indispensable to include top-mass effects in state-of-the-art simulations of Higgs-boson production 
through gluon fusion, but it also shows the high quality of previously employed approximations of these effects. Our newly developed 
$pp\to H$ \minnlo{}  generator at NNLO+PS in the full theory constitutes the most accurate tool to model signal events of the Higgs boson 
at the LHC, and we reckon that it will be a useful addition to ongoing and future experimental analyses.
With the development of this generator we pave the way for including also subleading quark-mass effects, such as those originating from bottom-quark 
and charm-quark loops, whose inclusion we leave to future work. Although more challenging, the calculation for light quarks can be approached
in a very similar manner.

\noindent {\bf Acknowledgements.}
We would like to thank Micha\l{} Czakon, Robert Harlander and Giulia Zanderighi for fruitful discussions.
We are also indebted to Robert Harlander and Giulia Zanderighi for comments on the manuscript.

\appendix

\setlength{\bibsep}{3.1pt}
\renewcommand{\em}{}
\normalem
\bibliographystyle{JHEP}
\bibliography{MiNNLO_H_full_theory}

\providecommand{\href}[2]{#2}\begingroup\raggedright\begin{thebibliography}{10}

\bibitem{ATLAS:2012yve}
{\scshape ATLAS} collaboration, \emph{{Observation of a new particle in the
  search for the Standard Model Higgs boson with the ATLAS detector at the
  LHC}}, \href{https://doi.org/10.1016/j.physletb.2012.08.020}{\emph{Phys.
  Lett. B} {\bfseries 716} (2012) 1}
  [\href{https://arxiv.org/abs/1207.7214}{{\ttfamily 1207.7214}}].

\bibitem{CMS:2012qbp}
{\scshape CMS} collaboration, \emph{{Observation of a New Boson at a Mass of
  125 GeV with the CMS Experiment at the LHC}},
  \href{https://doi.org/10.1016/j.physletb.2012.08.021}{\emph{Phys. Lett. B}
  {\bfseries 716} (2012) 30} [\href{https://arxiv.org/abs/1207.7235}{{\ttfamily
  1207.7235}}].

\bibitem{ATLAS:2022vkf}
{\scshape ATLAS} collaboration, \emph{{A detailed map of Higgs boson
  interactions by the ATLAS experiment ten years after the discovery}},
  \href{https://doi.org/10.1038/s41586-022-04893-w}{\emph{Nature} {\bfseries
  607} (2022) 52} [\href{https://arxiv.org/abs/2207.00092}{{\ttfamily
  2207.00092}}].

\bibitem{CMS:2022dwd}
{\scshape CMS} collaboration, \emph{{A portrait of the Higgs boson by the CMS
  experiment ten years after the discovery.}},
  \href{https://doi.org/10.1038/s41586-022-04892-x}{\emph{Nature} {\bfseries
  607} (2022) 60} [\href{https://arxiv.org/abs/2207.00043}{{\ttfamily
  2207.00043}}].

\bibitem{Harlander:2002wh}
R.V.~Harlander and W.B.~Kilgore, \emph{{Next-to-next-to-leading order Higgs
  production at hadron colliders}},
  \href{https://doi.org/10.1103/PhysRevLett.88.201801}{\emph{Phys.Rev.Lett.}
  {\bfseries 88} (2002) 201801}
  [\href{https://arxiv.org/abs/hep-ph/0201206}{{\ttfamily hep-ph/0201206}}].

\bibitem{Anastasiou:2002yz}
C.~Anastasiou and K.~Melnikov, \emph{{Higgs boson production at hadron
  colliders in NNLO QCD}},
  \href{https://doi.org/10.1016/S0550-3213(02)00837-4}{\emph{Nucl.Phys.}
  {\bfseries B646} (2002) 220}
  [\href{https://arxiv.org/abs/hep-ph/0207004}{{\ttfamily hep-ph/0207004}}].

\bibitem{Ravindran:2003um}
V.~Ravindran, J.~Smith and W.L.~van Neerven, \emph{{NNLO corrections to the
  total cross-section for Higgs boson production in hadron hadron collisions}},
  \href{https://doi.org/10.1016/S0550-3213(03)00457-7}{\emph{Nucl.Phys.}
  {\bfseries B665} (2003) 325}
  [\href{https://arxiv.org/abs/hep-ph/0302135}{{\ttfamily hep-ph/0302135}}].

\bibitem{Anastasiou:2015vya}
C.~Anastasiou, C.~Duhr, F.~Dulat, F.~Herzog and B.~Mistlberger, \emph{{Higgs
  Boson Gluon-Fusion Production in QCD at Three Loops}},
  \href{https://doi.org/10.1103/PhysRevLett.114.212001}{\emph{Phys. Rev. Lett.}
  {\bfseries 114} (2015) 212001}
  [\href{https://arxiv.org/abs/1503.06056}{{\ttfamily 1503.06056}}].

\bibitem{Mistlberger:2018etf}
B.~Mistlberger, \emph{{Higgs boson production at hadron colliders at N$^{3}$LO
  in QCD}}, \href{https://doi.org/10.1007/JHEP05(2018)028}{\emph{JHEP}
  {\bfseries 05} (2018) 028}
  [\href{https://arxiv.org/abs/1802.00833}{{\ttfamily 1802.00833}}].

\bibitem{Anastasiou:2004xq}
C.~Anastasiou, K.~Melnikov and F.~Petriello, \emph{{Higgs boson production at
  hadron colliders: Differential cross sections through next-to-next-to-leading
  order}}, \href{https://doi.org/10.1103/PhysRevLett.93.262002}{\emph{Phys.
  Rev. Lett.} {\bfseries 93} (2004) 262002}
  [\href{https://arxiv.org/abs/hep-ph/0409088}{{\ttfamily hep-ph/0409088}}].

\bibitem{Catani:2007vq}
S.~Catani and M.~Grazzini, \emph{{An NNLO subtraction formalism in hadron
  collisions and its application to Higgs boson production at the LHC}},
  \href{https://doi.org/10.1103/PhysRevLett.98.222002}{\emph{Phys. Rev. Lett.}
  {\bfseries 98} (2007) 222002}
  [\href{https://arxiv.org/abs/hep-ph/0703012}{{\ttfamily hep-ph/0703012}}].

\bibitem{Boughezal:2013uia}
R.~Boughezal, F.~Caola, K.~Melnikov, F.~Petriello and M.~Schulze, \emph{{Higgs
  boson production in association with a jet at next-to-next-to-leading order
  in perturbative QCD}},
  \href{https://doi.org/10.1007/JHEP06(2013)072}{\emph{JHEP} {\bfseries 06}
  (2013) 072} [\href{https://arxiv.org/abs/1302.6216}{{\ttfamily 1302.6216}}].

\bibitem{Chen:2014gva}
X.~Chen, T.~Gehrmann, E.W.N.~Glover and M.~Jaquier, \emph{{Precise QCD
  predictions for the production of Higgs + jet final states}},
  \href{https://doi.org/10.1016/j.physletb.2014.11.021}{\emph{Phys. Lett.}
  {\bfseries B740} (2015) 147}
  [\href{https://arxiv.org/abs/1408.5325}{{\ttfamily 1408.5325}}].

\bibitem{Boughezal:2015dra}
R.~Boughezal, F.~Caola, K.~Melnikov, F.~Petriello and M.~Schulze, \emph{{Higgs
  boson production in association with a jet at next-to-next-to-leading
  order}}, \href{https://doi.org/10.1103/PhysRevLett.115.082003}{\emph{Phys.
  Rev. Lett.} {\bfseries 115} (2015) 082003}
  [\href{https://arxiv.org/abs/1504.07922}{{\ttfamily 1504.07922}}].

\bibitem{Chen:2016zka}
X.~Chen, J.~Cruz-Martinez, T.~Gehrmann, E.W.N.~Glover and M.~Jaquier,
  \emph{{NNLO QCD corrections to Higgs boson production at large transverse
  momentum}}, \href{https://doi.org/10.1007/JHEP10(2016)066}{\emph{JHEP}
  {\bfseries 10} (2016) 066}
  [\href{https://arxiv.org/abs/1607.08817}{{\ttfamily 1607.08817}}].

\bibitem{Hamilton:2013fea}
K.~Hamilton, P.~Nason, E.~Re and G.~Zanderighi, \emph{{NNLOPS simulation of
  Higgs boson production}},
  \href{https://doi.org/10.1007/JHEP10(2013)222}{\emph{JHEP} {\bfseries 10}
  (2013) 222} [\href{https://arxiv.org/abs/1309.0017}{{\ttfamily 1309.0017}}].

\bibitem{Hoche:2014dla}
S.~Höche, Y.~Li and S.~Prestel, \emph{{Higgs-boson production through gluon
  fusion at NNLO QCD with parton showers}},
  \href{https://doi.org/10.1103/PhysRevD.90.054011}{\emph{Phys. Rev.}
  {\bfseries D90} (2014) 054011}
  [\href{https://arxiv.org/abs/1407.3773}{{\ttfamily 1407.3773}}].

\bibitem{Monni:2019whf}
P.F.~Monni, P.~Nason, E.~Re, M.~Wiesemann and G.~Zanderighi,
  \emph{{MiNNLO$_{\text{PS}}$: A new method to match NNLO QCD to parton
  showers}}, \href{https://doi.org/10.1007/JHEP05(2020)143}{\emph{JHEP}
  {\bfseries 05} (2020) 143}
  [\href{https://arxiv.org/abs/1908.06987}{{\ttfamily 1908.06987}}].

\bibitem{Monni:2020nks}
P.F.~Monni, E.~Re and M.~Wiesemann, \emph{{MiNNLO$_{\text {PS}}$: optimizing
  $2\rightarrow 1$ hadronic processes}},
  \href{https://doi.org/10.1140/epjc/s10052-020-08658-5}{\emph{Eur. Phys. J. C}
  {\bfseries 80} (2020) 1075}
  [\href{https://arxiv.org/abs/2006.04133}{{\ttfamily 2006.04133}}].

\bibitem{Cieri:2018oms}
L.~Cieri, X.~Chen, T.~Gehrmann, E.W.N.~Glover and A.~Huss, \emph{{Higgs boson
  production at the LHC using the $q_T$ subtraction formalism at N$^3$LO QCD}},
  \href{https://doi.org/10.1007/JHEP02(2019)096}{\emph{JHEP} {\bfseries 02}
  (2019) 096} [\href{https://arxiv.org/abs/1807.11501}{{\ttfamily
  1807.11501}}].

\bibitem{Dulat:2018bfe}
F.~Dulat, B.~Mistlberger and A.~Pelloni, \emph{{Precision predictions at
  N$^3$LO for the Higgs boson rapidity distribution at the LHC}},
  \href{https://doi.org/10.1103/PhysRevD.99.034004}{\emph{Phys. Rev.}
  {\bfseries D99} (2019) 034004}
  [\href{https://arxiv.org/abs/1810.09462}{{\ttfamily 1810.09462}}].

\bibitem{Chen:2021isd}
X.~Chen, T.~Gehrmann, E.W.N.~Glover, A.~Huss, B.~Mistlberger and A.~Pelloni,
  \emph{{Fully Differential Higgs Boson Production to Third Order in QCD}},
  \href{https://doi.org/10.1103/PhysRevLett.127.072002}{\emph{Phys. Rev. Lett.}
  {\bfseries 127} (2021) 072002}
  [\href{https://arxiv.org/abs/2102.07607}{{\ttfamily 2102.07607}}].

\bibitem{Billis:2021ecs}
G.~Billis, B.~Dehnadi, M.A.~Ebert, J.K.L.~Michel and F.J.~Tackmann,
  \emph{{Higgs pT Spectrum and Total Cross Section with Fiducial Cuts at Third
  Resummed and Fixed Order in QCD}},
  \href{https://doi.org/10.1103/PhysRevLett.127.072001}{\emph{Phys. Rev. Lett.}
  {\bfseries 127} (2021) 072001}
  [\href{https://arxiv.org/abs/2102.08039}{{\ttfamily 2102.08039}}].

\bibitem{Marzani:2008az}
S.~Marzani, R.D.~Ball, V.~Del~Duca, S.~Forte and A.~Vicini, \emph{{Higgs
  production via gluon-gluon fusion with finite top mass beyond next-to-leading
  order}}, \href{https://doi.org/10.1016/j.nuclphysb.2008.03.016}{\emph{Nucl.
  Phys.} {\bfseries B800} (2008) 127}
  [\href{https://arxiv.org/abs/0801.2544}{{\ttfamily 0801.2544}}].

\bibitem{Harlander:2009mq}
R.V.~Harlander and K.J.~Ozeren, \emph{{Finite top mass effects for hadronic
  Higgs production at next-to-next-to-leading order}},
  \href{https://doi.org/10.1088/1126-6708/2009/11/088}{\emph{JHEP} {\bfseries
  11} (2009) 088} [\href{https://arxiv.org/abs/0909.3420}{{\ttfamily
  0909.3420}}].

\bibitem{Harlander:2009my}
R.V.~Harlander, H.~Mantler, S.~Marzani and K.J.~Ozeren, \emph{{Higgs production
  in gluon fusion at next-to-next-to-leading order QCD for finite top mass}},
  \href{https://doi.org/10.1140/epjc/s10052-010-1258-x}{\emph{Eur. Phys. J.}
  {\bfseries C66} (2010) 359}
  [\href{https://arxiv.org/abs/0912.2104}{{\ttfamily 0912.2104}}].

\bibitem{Pak:2009dg}
A.~Pak, M.~Rogal and M.~Steinhauser, \emph{{Finite top quark mass effects in
  NNLO Higgs boson production at LHC}},
  \href{https://doi.org/10.1007/JHEP02(2010)025}{\emph{JHEP} {\bfseries 02}
  (2010) 025} [\href{https://arxiv.org/abs/0911.4662}{{\ttfamily 0911.4662}}].

\bibitem{Pak:2011hs}
A.~Pak, M.~Rogal and M.~Steinhauser, \emph{{Production of scalar and
  pseudo-scalar Higgs bosons to next-to-next-to-leading order at hadron
  colliders}}, \href{https://doi.org/10.1007/JHEP09(2011)088}{\emph{JHEP}
  {\bfseries 09} (2011) 088} [\href{https://arxiv.org/abs/1107.3391}{{\ttfamily
  1107.3391}}].

\bibitem{Spira:1995rr}
M.~Spira, A.~Djouadi, D.~Graudenz and P.M.~Zerwas, \emph{{Higgs boson
  production at the LHC}},
  \href{https://doi.org/10.1016/0550-3213(95)00379-7}{\emph{Nucl. Phys. B}
  {\bfseries 453} (1995) 17}
  [\href{https://arxiv.org/abs/hep-ph/9504378}{{\ttfamily hep-ph/9504378}}].

\bibitem{Harlander:2005rq}
R.~Harlander and P.~Kant, \emph{{Higgs production and decay: Analytic results
  at next-to-leading order QCD}},
  \href{https://doi.org/10.1088/1126-6708/2005/12/015}{\emph{JHEP} {\bfseries
  12} (2005) 015} [\href{https://arxiv.org/abs/hep-ph/0509189}{{\ttfamily
  hep-ph/0509189}}].

\bibitem{Bagnaschi:2011tu}
E.~Bagnaschi, G.~Degrassi, P.~Slavich and A.~Vicini, \emph{{Higgs production
  via gluon fusion in the POWHEG approach in the SM and in the MSSM}},
  \href{https://doi.org/10.1007/JHEP02(2012)088}{\emph{JHEP} {\bfseries 02}
  (2012) 088} [\href{https://arxiv.org/abs/1111.2854}{{\ttfamily 1111.2854}}].

\bibitem{Harlander:2012hf}
R.V.~Harlander, T.~Neumann, K.J.~Ozeren and M.~Wiesemann, \emph{{Top-mass
  effects in differential Higgs production through gluon fusion at order
  $\alpha_s^4$}}, \href{https://doi.org/10.1007/JHEP08(2012)139}{\emph{JHEP}
  {\bfseries 08} (2012) 139} [\href{https://arxiv.org/abs/1206.0157}{{\ttfamily
  1206.0157}}].

\bibitem{Mantler:2012bj}
H.~Mantler and M.~Wiesemann, \emph{{Top- and bottom-mass effects in hadronic
  Higgs production at small transverse momenta through LO+NLL}},
  \href{https://doi.org/10.1140/epjc/s10052-013-2467-x}{\emph{Eur. Phys. J. C}
  {\bfseries 73} (2013) 2467}
  [\href{https://arxiv.org/abs/1210.8263}{{\ttfamily 1210.8263}}].

\bibitem{Banfi:2013eda}
A.~Banfi, P.F.~Monni and G.~Zanderighi, \emph{{Quark masses in Higgs production
  with a jet veto}}, \href{https://doi.org/10.1007/JHEP01(2014)097}{\emph{JHEP}
  {\bfseries 01} (2014) 097} [\href{https://arxiv.org/abs/1308.4634}{{\ttfamily
  1308.4634}}].

\bibitem{Grazzini:2013mca}
M.~Grazzini and H.~Sargsyan, \emph{{Heavy-quark mass effects in Higgs boson
  production at the LHC}},
  \href{https://doi.org/10.1007/JHEP09(2013)129}{\emph{JHEP} {\bfseries 09}
  (2013) 129} [\href{https://arxiv.org/abs/1306.4581}{{\ttfamily 1306.4581}}].

\bibitem{Harlander:2014uea}
R.V.~Harlander, H.~Mantler and M.~Wiesemann, \emph{{Transverse momentum
  resummation for Higgs production via gluon fusion in the MSSM}},
  \href{https://doi.org/10.1007/JHEP11(2014)116}{\emph{JHEP} {\bfseries 1411}
  (2014) 116} [\href{https://arxiv.org/abs/1409.0531}{{\ttfamily 1409.0531}}].

\bibitem{Neumann:2014nha}
T.~Neumann and M.~Wiesemann, \emph{{Finite top-mass effects in gluon-induced
  Higgs production with a jet-veto at NNLO}},
  \href{https://doi.org/10.1007/JHEP11(2014)150}{\emph{JHEP} {\bfseries 11}
  (2014) 150} [\href{https://arxiv.org/abs/1408.6836}{{\ttfamily 1408.6836}}].

\bibitem{Hamilton:2015nsa}
K.~Hamilton, P.~Nason and G.~Zanderighi, \emph{{Finite quark-mass effects in
  the NNLOPS POWHEG+MiNLO Higgs generator}},
  \href{https://doi.org/10.1007/JHEP05(2015)140}{\emph{JHEP} {\bfseries 05}
  (2015) 140} [\href{https://arxiv.org/abs/1501.04637}{{\ttfamily
  1501.04637}}].

\bibitem{Mantler:2015vba}
H.~Mantler and M.~Wiesemann, \emph{{Hadronic Higgs production through NLO $+$
  PS in the SM, the 2HDM and the MSSM}},
  \href{https://doi.org/10.1140/epjc/s10052-015-3462-1}{\emph{Eur. Phys. J. C}
  {\bfseries 75} (2015) 257}
  [\href{https://arxiv.org/abs/1504.06625}{{\ttfamily 1504.06625}}].

\bibitem{Bagnaschi:2015qta}
E.~Bagnaschi and A.~Vicini, \emph{{The Higgs transverse momentum distribution
  in gluon fusion as a multiscale problem}},
  \href{https://doi.org/10.1007/JHEP01(2016)056}{\emph{JHEP} {\bfseries 01}
  (2016) 056} [\href{https://arxiv.org/abs/1505.00735}{{\ttfamily
  1505.00735}}].

\bibitem{Bagnaschi:2015bop}
E.~Bagnaschi, R.V.~Harlander, H.~Mantler, A.~Vicini and M.~Wiesemann,
  \emph{{Resummation ambiguities in the Higgs transverse-momentum spectrum in
  the Standard Model and beyond}},
  \href{https://doi.org/10.1007/JHEP01(2016)090}{\emph{JHEP} {\bfseries 01}
  (2016) 090} [\href{https://arxiv.org/abs/1510.08850}{{\ttfamily
  1510.08850}}].

\bibitem{Frederix:2016cnl}
R.~Frederix, S.~Frixione, E.~Vryonidou and M.~Wiesemann, \emph{{Heavy-quark
  mass effects in Higgs plus jets production}},
  \href{https://doi.org/10.1007/JHEP08(2016)006}{\emph{JHEP} {\bfseries 08}
  (2016) 006} [\href{https://arxiv.org/abs/1604.03017}{{\ttfamily
  1604.03017}}].

\bibitem{Neumann:2016dny}
T.~Neumann and C.~Williams, \emph{{The Higgs boson at high $p_T$}},
  \href{https://doi.org/10.1103/PhysRevD.95.014004}{\emph{Phys. Rev. D}
  {\bfseries 95} (2017) 014004}
  [\href{https://arxiv.org/abs/1609.00367}{{\ttfamily 1609.00367}}].

\bibitem{Melnikov:2016emg}
K.~Melnikov and A.~Penin, \emph{{On the light quark mass effects in Higgs boson
  production in gluon fusion}},
  \href{https://doi.org/10.1007/JHEP05(2016)172}{\emph{JHEP} {\bfseries 05}
  (2016) 172} [\href{https://arxiv.org/abs/1602.09020}{{\ttfamily
  1602.09020}}].

\bibitem{Caola:2016upw}
F.~Caola, S.~Forte, S.~Marzani, C.~Muselli and G.~Vita, \emph{{The Higgs
  transverse momentum spectrum with finite quark masses beyond leading order}},
  \href{https://doi.org/10.1007/JHEP08(2016)150}{\emph{JHEP} {\bfseries 08}
  (2016) 150} [\href{https://arxiv.org/abs/1606.04100}{{\ttfamily
  1606.04100}}].

\bibitem{Lindert:2017pky}
J.M.~Lindert, K.~Melnikov, L.~Tancredi and C.~Wever, \emph{{Top-bottom
  interference effects in Higgs plus jet production at the LHC}},
  \href{https://doi.org/10.1103/PhysRevLett.118.252002}{\emph{Phys. Rev. Lett.}
  {\bfseries 118} (2017) 252002}
  [\href{https://arxiv.org/abs/1703.03886}{{\ttfamily 1703.03886}}].

\bibitem{Liu:2017vkm}
T.~Liu and A.A.~Penin, \emph{{High-Energy Limit of QCD beyond the Sudakov
  Approximation}},
  \href{https://doi.org/10.1103/PhysRevLett.119.262001}{\emph{Phys. Rev. Lett.}
  {\bfseries 119} (2017) 262001}
  [\href{https://arxiv.org/abs/1709.01092}{{\ttfamily 1709.01092}}].

\bibitem{Jones:2018hbb}
S.P.~Jones, M.~Kerner and G.~Luisoni, \emph{{Next-to-Leading-Order QCD
  Corrections to Higgs Boson Plus Jet Production with Full Top-Quark Mass
  Dependence}},
  \href{https://doi.org/10.1103/PhysRevLett.120.162001}{\emph{Phys. Rev. Lett.}
  {\bfseries 120} (2018) 162001}
  [\href{https://arxiv.org/abs/1802.00349}{{\ttfamily 1802.00349}}].

\bibitem{Caola:2018zye}
F.~Caola, J.M.~Lindert, K.~Melnikov, P.F.~Monni, L.~Tancredi and C.~Wever,
  \emph{{Bottom-quark effects in Higgs production at intermediate transverse
  momentum}}, \href{https://doi.org/10.1007/JHEP09(2018)035}{\emph{JHEP}
  {\bfseries 09} (2018) 035}
  [\href{https://arxiv.org/abs/1804.07632}{{\ttfamily 1804.07632}}].

\bibitem{Chen:2021azt}
X.~Chen, A.~Huss, S.P.~Jones, M.~Kerner, J.N.~Lang, J.M.~Lindert et~al.,
  \emph{{Top-quark mass effects in H+jet and H+2 jets production}},
  \href{https://doi.org/10.1007/JHEP03(2022)096}{\emph{JHEP} {\bfseries 03}
  (2022) 096} [\href{https://arxiv.org/abs/2110.06953}{{\ttfamily
  2110.06953}}].

\bibitem{Czakon:2021yub}
M.~Czakon, R.V.~Harlander, J.~Klappert and M.~Niggetiedt, \emph{{Exact
  Top-Quark Mass Dependence in Hadronic Higgs Production}},
  \href{https://doi.org/10.1103/PhysRevLett.127.162002}{\emph{Phys. Rev. Lett.}
  {\bfseries 127} (2021) 162002}
  [\href{https://arxiv.org/abs/2105.04436}{{\ttfamily 2105.04436}}].

\bibitem{Czakon:2023kqm}
M.~Czakon, F.~Eschment, M.~Niggetiedt, R.~Poncelet and T.~Schellenberger,
  \emph{{Top-Bottom Interference Contribution to Fully Inclusive Higgs
  Production}},
  \href{https://doi.org/10.1103/PhysRevLett.132.211902}{\emph{Phys. Rev. Lett.}
  {\bfseries 132} (2024) 211902}
  [\href{https://arxiv.org/abs/2312.09896}{{\ttfamily 2312.09896}}].

\bibitem{Jezo:2015aia}
T.~Je\v{z}o and P.~Nason, \emph{{On the Treatment of Resonances in
  Next-to-Leading Order Calculations Matched to a Parton Shower}},
  \href{https://doi.org/10.1007/JHEP12(2015)065}{\emph{JHEP} {\bfseries 12}
  (2015) 065} [\href{https://arxiv.org/abs/1509.09071}{{\ttfamily
  1509.09071}}].

\bibitem{Cascioli:2011va}
F.~Cascioli, P.~Maierh\"ofer and S.~Pozzorini, \emph{{Scattering Amplitudes
  with Open Loops}},
  \href{https://doi.org/10.1103/PhysRevLett.108.111601}{\emph{Phys. Rev. Lett.}
  {\bfseries 108} (2012) 111601}
  [\href{https://arxiv.org/abs/1111.5206}{{\ttfamily 1111.5206}}].

\bibitem{Buccioni:2017yxi}
F.~Buccioni, S.~Pozzorini and M.~Zoller, \emph{{On-the-fly reduction of open
  loops}}, \href{https://doi.org/10.1140/epjc/s10052-018-5562-1}{\emph{Eur.
  Phys. J.} {\bfseries C78} (2018) 70}
  [\href{https://arxiv.org/abs/1710.11452}{{\ttfamily 1710.11452}}].

\bibitem{Buccioni:2019sur}
F.~Buccioni, J.-N.~Lang, J.M.~Lindert, P.~Maierh{\"o}fer, S.~Pozzorini,
  H.~Zhang et~al., \emph{{OpenLoops 2}},
  \href{https://doi.org/10.1140/epjc/s10052-019-7306-2}{\emph{Eur. Phys. J. C}
  {\bfseries 79} (2019) 866}
  [\href{https://arxiv.org/abs/1907.13071}{{\ttfamily 1907.13071}}].

\bibitem{Jezo:2016ujg}
T.~Je\v{z}o, J.M.~Lindert, P.~Nason, C.~Oleari and S.~Pozzorini, \emph{{An
  NLO+PS generator for $t\bar{t}$ and $Wt$ production and decay including
  non-resonant and interference effects}},
  \href{https://doi.org/10.1140/epjc/s10052-016-4538-2}{\emph{Eur. Phys. J. C}
  {\bfseries 76} (2016) 691}
  [\href{https://arxiv.org/abs/1607.04538}{{\ttfamily 1607.04538}}].

\bibitem{Czakon:2020vql}
M.L.~Czakon and M.~Niggetiedt, \emph{{Exact quark-mass dependence of the
  Higgs-gluon form factor at three loops in QCD}},
  \href{https://doi.org/10.1007/JHEP05(2020)149}{\emph{JHEP} {\bfseries 05}
  (2020) 149} [\href{https://arxiv.org/abs/2001.03008}{{\ttfamily
  2001.03008}}].

\bibitem{Maierhofer:2017gsa}
P.~Maierh\"ofer, J.~Usovitsch and P.~Uwer, \emph{{Kira\textemdash{}A Feynman
  integral reduction program}},
  \href{https://doi.org/10.1016/j.cpc.2018.04.012}{\emph{Comput. Phys. Commun.}
  {\bfseries 230} (2018) 99}
  [\href{https://arxiv.org/abs/1705.05610}{{\ttfamily 1705.05610}}].

\bibitem{Maierhofer:2018gpa}
P.~Maierh\"ofer and J.~Usovitsch, \emph{{Kira 1.2 Release Notes}},
  \href{https://arxiv.org/abs/1812.01491}{{\ttfamily 1812.01491}}.

\bibitem{Klappert:2020nbg}
J.~Klappert, F.~Lange, P.~Maierh\"ofer and J.~Usovitsch, \emph{{Integral
  reduction with Kira 2.0 and finite field methods}},
  \href{https://doi.org/10.1016/j.cpc.2021.108024}{\emph{Comput. Phys. Commun.}
  {\bfseries 266} (2021) 108024}
  [\href{https://arxiv.org/abs/2008.06494}{{\ttfamily 2008.06494}}].

\bibitem{Klappert:2019emp}
J.~Klappert and F.~Lange, \emph{{Reconstructing rational functions with
  FireFly}}, \href{https://doi.org/10.1016/j.cpc.2019.106951}{\emph{Comput.
  Phys. Commun.} {\bfseries 247} (2020) 106951}
  [\href{https://arxiv.org/abs/1904.00009}{{\ttfamily 1904.00009}}].

\bibitem{Klappert:2020aqs}
J.~Klappert, S.Y.~Klein and F.~Lange, \emph{{Interpolation of dense and sparse
  rational functions and other improvements in FireFly}},
  \href{https://doi.org/10.1016/j.cpc.2021.107968}{\emph{Comput. Phys. Commun.}
  {\bfseries 264} (2021) 107968}
  [\href{https://arxiv.org/abs/2004.01463}{{\ttfamily 2004.01463}}].

\bibitem{Kotikov:1990kg}
A.V.~Kotikov, \emph{{Differential equations method: New technique for massive
  Feynman diagrams calculation}},
  \href{https://doi.org/10.1016/0370-2693(91)90413-K}{\emph{Phys. Lett. B}
  {\bfseries 254} (1991) 158}.

\bibitem{Kotikov:1991pm}
A.V.~Kotikov, \emph{{Differential equation method: The Calculation of N point
  Feynman diagrams}},
  \href{https://doi.org/10.1016/0370-2693(91)90536-Y}{\emph{Phys. Lett. B}
  {\bfseries 267} (1991) 123}.

\bibitem{Kotikov:1991hm}
A.V.~Kotikov, \emph{{Differential equations method: The Calculation of vertex
  type Feynman diagrams}},
  \href{https://doi.org/10.1016/0370-2693(91)90834-D}{\emph{Phys. Lett. B}
  {\bfseries 259} (1991) 314}.

\bibitem{Remiddi:1997ny}
E.~Remiddi, \emph{{Differential equations for Feynman graph amplitudes}},
  \href{https://doi.org/10.1007/BF03185566}{\emph{Nuovo Cim. A} {\bfseries 110}
  (1997) 1435} [\href{https://arxiv.org/abs/hep-th/9711188}{{\ttfamily
  hep-th/9711188}}].

\bibitem{Gorishnii:1989dd}
S.G.~Gorishnii, \emph{{Construction of Operator Expansions and Effective
  Theories in the Ms Scheme}},
  \href{https://doi.org/10.1016/0550-3213(89)90622-6}{\emph{Nucl. Phys. B}
  {\bfseries 319} (1989) 633}.

\bibitem{Smirnov:1990rz}
V.A.~Smirnov, \emph{{Asymptotic expansions in limits of large momenta and
  masses}}, \href{https://doi.org/10.1007/BF02102092}{\emph{Commun. Math.
  Phys.} {\bfseries 134} (1990) 109}.

\bibitem{Smirnov:1994tg}
V.A.~Smirnov, \emph{{Asymptotic expansions in momenta and masses and
  calculation of Feynman diagrams}},
  \href{https://doi.org/10.1142/S0217732395001617}{\emph{Mod. Phys. Lett. A}
  {\bfseries 10} (1995) 1485}
  [\href{https://arxiv.org/abs/hep-th/9412063}{{\ttfamily hep-th/9412063}}].

\bibitem{Smirnov:2002pj}
V.A.~Smirnov, \emph{{Applied asymptotic expansions in momenta and masses}},
  {\emph{Springer Tracts Mod. Phys.} {\bfseries 177} (2002) 1}.

\bibitem{Bonciani:2016qxi}
R.~Bonciani, V.~Del~Duca, H.~Frellesvig, J.M.~Henn, F.~Moriello and
  V.A.~Smirnov, \emph{{Two-loop planar master integrals for Higgs$\to 3$
  partons with full heavy-quark mass dependence}},
  \href{https://doi.org/10.1007/JHEP12(2016)096}{\emph{JHEP} {\bfseries 12}
  (2016) 096} [\href{https://arxiv.org/abs/1609.06685}{{\ttfamily
  1609.06685}}].

\bibitem{Bonciani:2019jyb}
R.~Bonciani, V.~Del~Duca, H.~Frellesvig, J.M.~Henn, M.~Hidding, L.~Maestri
  et~al., \emph{{Evaluating a family of two-loop non-planar master integrals
  for Higgs + jet production with full heavy-quark mass dependence}},
  \href{https://doi.org/10.1007/JHEP01(2020)132}{\emph{JHEP} {\bfseries 01}
  (2020) 132} [\href{https://arxiv.org/abs/1907.13156}{{\ttfamily
  1907.13156}}].

\bibitem{Frellesvig:2019byn}
H.~Frellesvig, M.~Hidding, L.~Maestri, F.~Moriello and G.~Salvatori, \emph{{The
  complete set of two-loop master integrals for Higgs + jet production in
  QCD}}, \href{https://doi.org/10.1007/JHEP06(2020)093}{\emph{JHEP} {\bfseries
  06} (2020) 093} [\href{https://arxiv.org/abs/1911.06308}{{\ttfamily
  1911.06308}}].

\bibitem{Niggetiedt:2023uyk}
M.~Niggetiedt and J.~Usovitsch, \emph{{The Higgs-gluon form factor at three
  loops in QCD with three mass scales}},
  \href{https://doi.org/10.1007/JHEP02(2024)087}{\emph{JHEP} {\bfseries 02}
  (2024) 087} [\href{https://arxiv.org/abs/2312.05297}{{\ttfamily
  2312.05297}}].

\bibitem{Becher:2009cu}
T.~Becher and M.~Neubert, \emph{{Infrared singularities of scattering
  amplitudes in perturbative QCD}},
  \href{https://doi.org/10.1103/PhysRevLett.102.162001}{\emph{Phys. Rev. Lett.}
  {\bfseries 102} (2009) 162001}
  [\href{https://arxiv.org/abs/0901.0722}{{\ttfamily 0901.0722}}].

\bibitem{Becher:2009qa}
T.~Becher and M.~Neubert, \emph{{On the Structure of Infrared Singularities of
  Gauge-Theory Amplitudes}},
  \href{https://doi.org/10.1088/1126-6708/2009/06/081}{\emph{JHEP} {\bfseries
  06} (2009) 081} [\href{https://arxiv.org/abs/0903.1126}{{\ttfamily
  0903.1126}}].

\bibitem{Lombardi:2020wju}
D.~Lombardi, M.~Wiesemann and G.~Zanderighi, \emph{{Advancing M\i{}NNLO$_{PS}$
  to diboson processes: Z\ensuremath{\gamma} production at NNLO+PS}},
  \href{https://doi.org/10.1007/JHEP06(2021)095}{\emph{JHEP} {\bfseries 06}
  (2021) 095} [\href{https://arxiv.org/abs/2010.10478}{{\ttfamily
  2010.10478}}].

\bibitem{Lombardi:2021rvg}
D.~Lombardi, M.~Wiesemann and G.~Zanderighi, \emph{{W$^{+}$W$^{-}$ production
  at NNLO+PS with MiNNLO$_{\rm PS}$}},
  \href{https://doi.org/10.1007/JHEP11(2021)230}{\emph{JHEP} {\bfseries 11}
  (2021) 230} [\href{https://arxiv.org/abs/2103.12077}{{\ttfamily
  2103.12077}}].

\bibitem{Buonocore:2021fnj}
L.~Buonocore, G.~Koole, D.~Lombardi, L.~Rottoli, M.~Wiesemann and
  G.~Zanderighi, \emph{{ZZ production at nNNLO+PS with MiNNLO$_{PS}$}},
  \href{https://doi.org/10.1007/JHEP01(2022)072}{\emph{JHEP} {\bfseries 01}
  (2022) 072} [\href{https://arxiv.org/abs/2108.05337}{{\ttfamily
  2108.05337}}].

\bibitem{Lombardi:2021wug}
D.~Lombardi, M.~Wiesemann and G.~Zanderighi, \emph{{Anomalous couplings in
  Z$\gamma$ events at NNLO+PS and improving $\nu\bar{\nu}\gamma$ backgrounds in
  dark-matter searches}},
  \href{https://doi.org/10.1016/j.physletb.2021.136846}{\emph{Phys. Lett. B}
  {\bfseries 824} (2022) 136846}
  [\href{https://arxiv.org/abs/2108.11315}{{\ttfamily 2108.11315}}].

\bibitem{Zanoli:2021iyp}
S.~Zanoli, M.~Chiesa, E.~Re, M.~Wiesemann and G.~Zanderighi,
  \emph{{Next-to-next-to-leading order event generation for VH production with
  $H\to b\overline{b} $ decay}},
  \href{https://doi.org/10.1007/JHEP07(2022)008}{\emph{JHEP} {\bfseries 07}
  (2022) 008} [\href{https://arxiv.org/abs/2112.04168}{{\ttfamily
  2112.04168}}].

\bibitem{Gavardi:2022ixt}
A.~Gavardi, C.~Oleari and E.~Re, \emph{{NNLO+PS Monte Carlo simulation of
  photon pair production with MiNNLOPS}},
  \href{https://arxiv.org/abs/2204.12602}{{\ttfamily 2204.12602}}.

\bibitem{Haisch:2022nwz}
U.~Haisch, D.J.~Scott, M.~Wiesemann, G.~Zanderighi and S.~Zanoli, \emph{{NNLO
  event generation for $ pp\to Zh\to
  {\mathrm{\ell}}^{+}{\mathrm{\ell}}^{-}b\overline{b} $ production in the SM
  effective field theory}},
  \href{https://doi.org/10.1007/JHEP07(2022)054}{\emph{JHEP} {\bfseries 07}
  (2022) 054} [\href{https://arxiv.org/abs/2204.00663}{{\ttfamily
  2204.00663}}].

\bibitem{Lindert:2022qdd}
J.M.~Lindert, D.~Lombardi, M.~Wiesemann, G.~Zanderighi and S.~Zanoli,
  \emph{{W$^{±}$Z production at NNLO QCD and NLO EW matched to parton showers
  with MiNNLO$_{PS}$}},
  \href{https://doi.org/10.1007/JHEP11(2022)036}{\emph{JHEP} {\bfseries 11}
  (2022) 036} [\href{https://arxiv.org/abs/2208.12660}{{\ttfamily
  2208.12660}}].

\bibitem{Biello:2024vdh}
C.~Biello, A.~Sankar, M.~Wiesemann and G.~Zanderighi, \emph{{NNLO+PS
  predictions for Higgs production through bottom-quark annihilation with
  MiNNLO$_{\text{PS}}$}},
  \href{https://doi.org/10.1140/epjc/s10052-024-12845-z}{\emph{Eur. Phys. J. C}
  {\bfseries 84} (2024) 479}
  [\href{https://arxiv.org/abs/2402.04025}{{\ttfamily 2402.04025}}].

\bibitem{Mazzitelli:2020jio}
J.~Mazzitelli, P.F.~Monni, P.~Nason, E.~Re, M.~Wiesemann and G.~Zanderighi,
  \emph{{Next-to-Next-to-Leading Order Event Generation for Top-Quark Pair
  Production}},
  \href{https://doi.org/10.1103/PhysRevLett.127.062001}{\emph{Phys. Rev. Lett.}
  {\bfseries 127} (2021) 062001}
  [\href{https://arxiv.org/abs/2012.14267}{{\ttfamily 2012.14267}}].

\bibitem{Mazzitelli:2021mmm}
J.~Mazzitelli, P.F.~Monni, P.~Nason, E.~Re, M.~Wiesemann and G.~Zanderighi,
  \emph{{Top-pair production at the LHC with MINNLO$_{PS}$}},
  \href{https://doi.org/10.1007/JHEP04(2022)079}{\emph{JHEP} {\bfseries 04}
  (2022) 079} [\href{https://arxiv.org/abs/2112.12135}{{\ttfamily
  2112.12135}}].

\bibitem{Mazzitelli:2023znt}
J.~Mazzitelli, A.~Ratti, M.~Wiesemann and G.~Zanderighi, \emph{{B-hadron
  production at the LHC from bottom-quark pair production at NNLO+PS}},
  \href{https://doi.org/10.1016/j.physletb.2023.137991}{\emph{Phys. Lett. B}
  {\bfseries 843} (2023) 137991}
  [\href{https://arxiv.org/abs/2302.01645}{{\ttfamily 2302.01645}}].

\bibitem{Mazzitelli:2024ura}
J.~Mazzitelli, V.~Sotnikov and M.~Wiesemann, \emph{{Next-to-next-to-leading
  order event generation for Z-boson production in association with a
  bottom-quark pair}},  \href{https://arxiv.org/abs/2404.08598}{{\ttfamily
  2404.08598}}.

\bibitem{Nason:2004rx}
P.~Nason, \emph{{A New method for combining NLO QCD with shower Monte Carlo
  algorithms}},
  \href{https://doi.org/10.1088/1126-6708/2004/11/040}{\emph{JHEP} {\bfseries
  11} (2004) 040} [\href{https://arxiv.org/abs/hep-ph/0409146}{{\ttfamily
  hep-ph/0409146}}].

\bibitem{Nason:2006hfa}
P.~Nason and G.~Ridolfi, \emph{{A Positive-weight next-to-leading-order Monte
  Carlo for Z pair hadroproduction}},
  \href{https://doi.org/10.1088/1126-6708/2006/08/077}{\emph{JHEP} {\bfseries
  08} (2006) 077} [\href{https://arxiv.org/abs/hep-ph/0606275}{{\ttfamily
  hep-ph/0606275}}].

\bibitem{Frixione:2007vw}
S.~Frixione, P.~Nason and C.~Oleari, \emph{{Matching NLO QCD computations with
  Parton Shower simulations: the POWHEG method}},
  \href{https://doi.org/10.1088/1126-6708/2007/11/070}{\emph{JHEP} {\bfseries
  11} (2007) 070} [\href{https://arxiv.org/abs/0709.2092}{{\ttfamily
  0709.2092}}].

\bibitem{Alioli:2010xd}
S.~Alioli, P.~Nason, C.~Oleari and E.~Re, \emph{{A general framework for
  implementing NLO calculations in shower Monte Carlo programs: the POWHEG
  BOX}}, \href{https://doi.org/10.1007/JHEP06(2010)043}{\emph{JHEP} {\bfseries
  06} (2010) 043} [\href{https://arxiv.org/abs/1002.2581}{{\ttfamily
  1002.2581}}].

\bibitem{Ball:2017nwa}
{\scshape NNPDF} collaboration, \emph{{Parton distributions from high-precision
  collider data}},
  \href{https://doi.org/10.1140/epjc/s10052-017-5199-5}{\emph{Eur. Phys. J.}
  {\bfseries C77} (2017) 663}
  [\href{https://arxiv.org/abs/1706.00428}{{\ttfamily 1706.00428}}].

\bibitem{Sjostrand:2014zea}
T.~Sjöstrand, S.~Ask, J.R.~Christiansen, R.~Corke, N.~Desai, P.~Ilten et~al.,
  \emph{{An Introduction to PYTHIA 8.2}},
  \href{https://doi.org/10.1016/j.cpc.2015.01.024}{\emph{Comput. Phys. Commun.}
  {\bfseries 191} (2015) 159}
  [\href{https://arxiv.org/abs/1410.3012}{{\ttfamily 1410.3012}}].

\bibitem{ATLAS:2022qef}
{\scshape ATLAS} collaboration, \emph{{Measurement of the total and
  differential Higgs boson production cross-sections at $ \sqrt{s} $ = 13 TeV
  with the ATLAS detector by combining the H \textrightarrow{}
  ZZ$^{*}$\textrightarrow{} 4\ensuremath{\ell} and H \textrightarrow{}
  \ensuremath{\gamma}\ensuremath{\gamma} decay channels}},
  \href{https://doi.org/10.1007/JHEP05(2023)028}{\emph{JHEP} {\bfseries 05}
  (2023) 028} [\href{https://arxiv.org/abs/2207.08615}{{\ttfamily
  2207.08615}}].

\bibitem{Salam:2021tbm}
G.P.~Salam and E.~Slade, \emph{{Cuts for two-body decays at colliders}},
  \href{https://doi.org/10.1007/JHEP11(2021)220}{\emph{JHEP} {\bfseries 11}
  (2021) 220} [\href{https://arxiv.org/abs/2106.08329}{{\ttfamily
  2106.08329}}].

\bibitem{Grazzini:2017mhc}
M.~Grazzini, S.~Kallweit and M.~Wiesemann, \emph{{Fully differential NNLO
  computations with MATRIX}},
  \href{https://doi.org/10.1140/epjc/s10052-018-5771-7}{\emph{Eur. Phys. J.}
  {\bfseries C78} (2018) 537}
  [\href{https://arxiv.org/abs/1711.06631}{{\ttfamily 1711.06631}}].

\end{thebibliography}\endgroup
\end{document}